\documentclass[arps,twocolumn,showpacs]{revtex4}

\usepackage{bm}
\usepackage{graphicx}
\usepackage{subfigure}
\input epsf

\newcommand\lsim{\mathrel{\rlap{\lower4pt\hbox{\hskip1pt$\sim$}}
        \raise1pt\hbox{$<$}}}
\newcommand\gsim{\mathrel{\rlap{\lower4pt\hbox{\hskip1pt$\sim$}}
        \raise1pt\hbox{$>$}}}

\def\thetaB{\mbox{\boldmath$\theta$}}
\def\thetagB{\mbox{\boldmath$\theta_g$}}
\def\dthetaB{\delta \mbox{\boldmath$\theta$}}

\begin{document}
\title{Anisotropic Magnification Distortion of the 3D \\
Galaxy Correlation: I. Real Space}

\author{Lam Hui$^{1,2}$, Enrique Gazta\~{n}aga$^{3}$ and Marilena LoVerde$^{1,2}$}

\affiliation{
$^{1}$Institute for Strings, Cosmology and Astroparticle Physics (ISCAP)\\
$^{2}$Department of Physics, Columbia University, New York, NY 10027\\
$^{3}$Institut de Ci\`encies de l'Espai, IEEC-CSIC, Campus UAB,
F. de Ci\`encies, Torre C5 par-2,  Barcelona 08193, Spain\\
lhui@astro.columbia.edu, gazta@ieec.uab.es, marilena@phys.columbia.edu
}
\date{\today}

\begin{abstract}
It has long been known that gravitational lensing, 
primarily via magnification bias, modifies the observed
galaxy (or quasar) clustering. Such discussions have largely focused on the 2D angular correlation
function. Here and in a companion paper (Paper II) we
explore how magnification bias distorts the 3D correlation function
and power spectrum, as first considered by
Matsubara (2000). 
The interesting point is: the distortion is anisotropic.
Magnification bias in general preferentially enhances the observed correlation
in the line-of-sight (LOS) orientation, especially on large scales.
For instance, at a LOS separation of $\sim 100$ Mpc/h, where the intrinsic 
galaxy-galaxy correlation is rather weak, the observed correlation can be enhanced 
by lensing by a factor of a few, even at a modest redshift of $z \sim 0.35$. 
This effect presents an interesting opportunity as well as challenge.
The opportunity: this lensing anisotropy is distinctive, making it possible to separately measure
the galaxy-galaxy, galaxy-magnification {\it and} magnification-magnification
correlations, without measuring galaxy shapes.  
The anisotropy is distinguishable from the well known distortion
due to peculiar motions, as will be discussed in Paper II.
The challenge: the magnification distortion
of the galaxy correlation must be accounted for in interpreting data as precision improves.
For instance, the $\sim 100$ Mpc/h baryon acoustic oscillation scale in the correlation function
is shifted by up to $\sim 3 \%$  in the LOS orientation, and up to
$\sim 0.6 \%$ in the monopole, depending on the galaxy bias, redshift and number count slope.
The corresponding shifts in the inferred Hubble parameter and angular
diameter distance, if ignored, could significantly bias measurements of
the dark energy equation of state.
Lastly, magnification distortion offers a plausible explanation
for the well known excess correlations seen in pencil beam surveys.
\end{abstract}

\pacs{98.80.-k; 98.80.Es; 98.65.Dx; 95.35.+d}


\maketitle

\section{Introduction}
\label{intro}
 
Since the pioneering work of \cite{TOG84}, it has been appreciated
that gravitational lensing modifies the spatial distribution of high redshift
objects, such as galaxies or quasars (henceforth, the term 
`galaxies' can be viewed as synonymous with quasars or any sample of survey objects).
Suppose for instance there is a large mass concentration between some galaxies and 
the observer. The observed number density of galaxies would decrease due to
the stretching of the apparent inter-galaxy spacing, but would increase due to 
the enhanced ability to see very faint galaxies that otherwise would have gone undetected.
The size of the net effect depends on the number count slope.
This effect is known as magnification bias.

The implications of
this effect for the observed galaxy angular correlation function were worked out
in \cite{J95,VFC97,kaiser98,MJV98}. It was 
pointed out by \cite{MJ98} that the magnification bias correction
can be isolated by measuring the angular cross-correlation between galaxies
at widely separated redshifts. This idea was subsequently realized in measurements
by \cite{EGmag03,ScrantonSDSS05}
through the cross-correlation of quasars and galaxies.
See \cite{menard,bhuv,JSS03} for some recent
theoretical work and references therein for earlier discussions of quasar-galaxy
associations. 
See also \cite{BTP95} on the early use of magnification bias to make mass maps of galaxy
clusters.

With the important exception of Matsubara (2000) \cite{matsubara00}
(to which we will return at the end of this section),
earlier papers have focused on the 2D angular correlation function.
Here and in Paper II \cite{3Dpaper2} of this series, we study 
the effect of magnification bias on the 3D
correlation function and power spectrum. 
It is not hard to convince oneself that lensing makes the 3D correlation
anisotropic. There are several effects at work, but the simplest one to
think about is the following. Correlation function is measured by
pair counts of galaxies. A pair of galaxies that are aligned along the 
line-of-sight (LOS) behave differently from a pair oriented transverse to
the LOS. In the former case, the closer galaxy can lens the background one.
The same does not happen in the transverse orientation. The net effect is
an anisotropic observed correlation. 
One might think such an effect must be small: after all, typical
LOS separations in clustering measurements are much smaller than the depth of surveys
i.e. the lens is located much closer to the source than to the observer.
However, one must remember that the lensing effect {\it grows} with the LOS
separation, whereas the intrinsic galaxy correlation generally {\it drops} with
separation. At a sufficiently large LOS separation, e.g. $\sim 100$ Mpc/h,
where the intrinsic galaxy correlation is rather weak, one should not
be too surprised that the lensing induced correlation can actually dominate.
As we will show, the magnification bias induced anisotropy has a 
distinctive shape:
\begin{eqnarray}
\label{scaling}
\xi_{\rm obs} (\delta\chi, \delta x_\perp)
= && \xi_{gg} (\sqrt{\delta\chi^2 + \delta x_\perp^2}) \\ \nonumber && 
+ f(\delta x_\perp) \delta\chi
+ g(\delta x_\perp)
\end{eqnarray}
where $\delta\chi$ and $\delta x_\perp$ are the LOS and transverse separations
respectively, $\xi_{\rm obs}$ is the observed correlation,
$\xi_{gg}$ is the intrinsic galaxy-galaxy correlation, 
$f\delta\chi$ is the galaxy-magnification correlation and
$g$ is the magnification-magnification correlation. Note that
$f$ and $g$ are functions of the transverse separation only.

This distinctive anisotropy pattern makes it in principle possible to 
completely separate the three different contributions from data: galaxy-galaxy,
galaxy-magnification and magnification-magnification correlations.
(We will discuss other sources of anisotropy, such as peculiar motions in Paper II.)
In a sense, the earlier work \cite{MJ98,EGmag03,ScrantonSDSS05,
menard,bhuv,JSS03} on angular correlation function
for galaxies (or galaxies and quasars) at widely separated redshifts
focused on one particular limit: a very large LOS separation $\delta\chi$ such that
the galaxy-magnification correlation $f\delta\chi$ dominates.
Studying more moderate LOS separations allows one to measure
the magnification-magnification correlation $g$, which is perhaps
of more theoretical interest since it relates directly to the mass.
The key is to use the full 3D information, i.e. exploit the
distinct dependence on $\delta\chi$
and $\delta x_\perp$ of each term in eq. (\ref{scaling}) to separately determine 
all three correlations.

Another important implication of eq. (\ref{scaling}) is that care must
be taken in interpreting galaxy clustering data.
For instance, future galaxy surveys hope to determine the baryon oscillation
scale to high precision \cite{eisenstein,2dFa,2dFb,hutsi,tegmark,baotheory,baoexp}.
Lensing induced shifts of the apparent baryon
oscillation scale in the LOS direction affects the inference on
the Hubble parameter $H(z)$, while shifts in the transverse direction affects
the inference on the angular diameter distance.
Recall that at a redshift of $1$, a $\sim 1 \%$ shift
in angular diameter distance, or a $\sim 1 \%$ shift in the
Hubble parameter, corresponds to a $\sim 5 \%$ shift in the dark
energy equation of state. 
This means that even small lensing corrections are
in principle a worry, 
not to mention potentially large corrections
at high redshifts or at particularly susceptible orientations, 
such as the LOS direction. 

The rest of the paper is organized as follows.
In \S \ref{distort}, we derive and numerically compute the
magnification bias distortion, magnification distortion in short,
of the observed galaxy clustering --
\S \ref{correlation} focuses on the overall anisotropy of the
 correlation function while \S \ref{sec:BAO} focuses on the
 baryon acoustic oscillations. In \S\ref{sec:interp} we present
useful order of magnitude estimates which offer deeper insights
into the numerical results of \S\ref{distort}.
 We conclude in \S \ref{discuss}, with a discussion of
longstanding puzzles posed by pencil beam surveys.
Appendix \ref{app:lens} contains a discussion of lensing
corrections to the observed correlation 
other than those due to magnification bias, and a generalization
of the magnification bias effect to account for more complicated
galaxy selection. 
Appendix \ref{app:lens2} contains a discussion of the higher order
Taylor expansion terms that are ignored in \S \ref{correlation}.

In this paper, we focus exclusively on the effects of magnification
distortion in configuration/real space. Paper II of this series
explores this effect in Fourier space.
Redshift space distortion due to peculiar motion 
and the Alcock-Paczynski effect will be discussed there as well.
In yet another paper, we study the effects of magnification bias on
the angular galaxy correlation function, focusing on the impacts
on features of the power spectrum \cite{2Dpaper}.

As this paper was being completed, two preprints appeared which discussed some
related issues. Vallinotto et al. \cite{scott} explored the impact of lensing, especially
magnification bias, on baryon oscillation measurements. They focused
on the 2 point correlation function where the 2 points are at exactly the same redshift.
They did not examine the full 3D correlation, in particular its anisotropy.
Their findings are consistent with ours for pair separations oriented
transverse to the LOS, and are more connected to our other paper on the
angular correlation function \cite{2Dpaper}. 
Wagner et al. 
\cite{steinmetz} examined the anisotropy of the 3D correlation that
is introduced by light cone effects. 

After this paper was initially circulated as a preprint, a
pioneering paper by Matsubara (2000) \cite{matsubara00}
was kindly brought to our attention, where he derived an expression for 3D correlation 
function in the presence of magnification bias as well as redshift and cosmological 
distortions.
In this paper, we have extended his analysis in a number of ways.
By showing the 3D correlation function in terms of the
LOS and transverse comoving separations (as opposed to
redshift and angular separations as in \cite{matsubara00}),
the anisotropy pattern is brought out explicitly.
We also emphasize the possibility to completely separate the
three different contributions to the observed correlation
function (see Fig. \ref{xi.fit} below). As an application, we show
how magnification distortion impacts baryon acoustic oscillation
measurements.
In Paper II, we also study the appearance of magnification
distortion in Fourier space (as well as redshift space), which 
turns out to have some important and interesting qualitative differences from the 
appearance in configuration/real space.

\section{Magnification Distortion}
\label{distort}

Magnification bias introduces a well-known correction to the
observed galaxy overdensity \cite{TOG84,J95,VFC97,MJV98,MJ98}:
\begin{eqnarray}
\label{start}
\delta_{\rm obs} = \delta_g + \delta_\mu
\end{eqnarray}
where the observed galaxy overdensity $\delta_{\rm obs}$, 
the original/intrinsic galaxy overdensity $\delta_g$ and the magnification bias
correction $\delta_\mu$ are all functions of
the galaxy position, specified by the radial comoving distance $\chi$
and the angular position $\thetaB$. 
The magnification bias correction is given by
\begin{eqnarray}
\label{deltamu}
\delta_\mu = (5s - 2) \kappa
\end{eqnarray}
where $\kappa$ is the lensing convergence:
\begin{eqnarray}
\kappa(\chi,\thetaB) = \int_0^\chi d\chi' {\chi'(\chi-\chi') \over \chi} \nabla^2_\perp \phi(\chi', \thetaB)
\end{eqnarray}
where $\phi$ is the gravitational potential, and
$\nabla^2_\perp$ is the 2D Laplacian in the transverse directions. We assume a flat universe -- 
generalization to an open or a closed universe is straightforward. 
The symbol $s$ stands for
\begin{eqnarray}
s = {d {\,\rm log}_{10} N(< m) \over dm}
\end{eqnarray}
where $N(< m)$ is the cumulative number counts for galaxies brighter
than magnitude $m$. This assumes the galaxy sample is defined by
a sharp faint-end cut-off. 
A broader definition of $s$ for a more general galaxy selection is given in Appendix \ref{app:lens}.

\begin{figure}[tb]
\centerline{\epsfxsize=9cm\epsffile{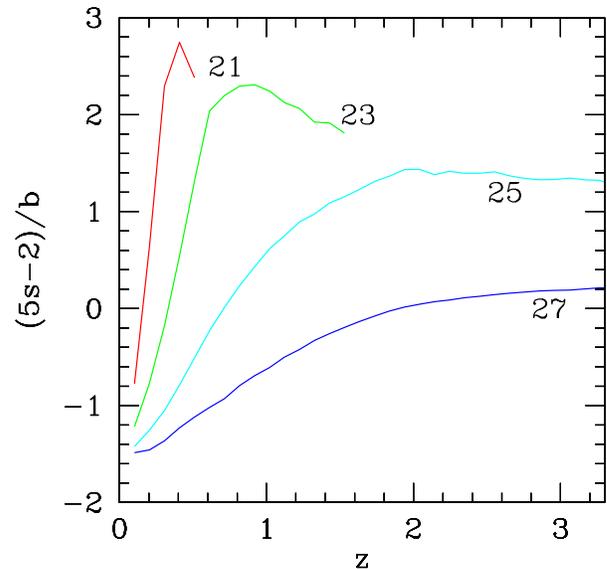}}
\caption{The ratio $(5s-2)/b$, where $s$ is the number count slope and 
$b$ is the galaxy bias, as a function of redshift $z$ for five different
galaxy samples. Each sample is defined by a B-band 
($4344$ angstroms) apparent magnitude cut-off
as shown for each curve. The curves should be viewed as illustrative
rather than definitive: only linear galaxy bias is accounted for here, and
the precise values for $s$ and $b$ depend on how the galaxies/quasars
are selected e.g. subject to color cuts and so on.
}
\label{b.s3}
\end{figure}

Defining the galaxy bias $b$ by $\delta_g = b \delta$, where $\delta$ is the mass overdensity,
eq. (\ref{start}) can be rewritten as
\begin{eqnarray}
\label{start2}
{\delta_{\rm obs} \over b} = \delta + {5s-2 \over b} \kappa
\end{eqnarray}
The relative importance of
the intrinsic clustering and the magnification bias correction is therefore
controlled by, among other things, the sample dependent ratio
$(5s-2)/b$. The observed redshift-dependent luminosity function \cite{gabasch}
can be used together with the halo model to estimate this ratio
(see \cite{LHGisw} for details).
Fig. \ref{b.s3} shows this as a function of redshift for five
different samples, each defined by a different B-band apparent magnitude cut-off.
This figure should be viewed as an illustration of the range of possibilities only.
The precise values of $s$ and $b$ depend sensitively on details of how the galaxy/quasar
sample is selected, for instance subject to color cuts and so on.
Unless otherwise stated, we adopt throughout this paper the value $(5s-2)/b = 1$ to illustrate
the effect of magnification bias on clustering measurements. 
For the correlation function or power spectrum, one can roughly scale
the magnification bias correction we obtain by $(5s-2)/b$ for galaxy redshift $\lsim 1.5$
and by $(5s-2)^2/b^2$ for redshift $\gsim 1.5$ (the former is dominated
by the galaxy-magnification cross-term while the latter is dominated by
the magnification-magnification term; see \S \ref{correlation}). Note also that
we assume a scale independent (linear) galaxy bias $b$. Nonlinear galaxy bias,
as we will see, is important in certain situations including, surprisingly,
some where large scale clustering measurements are involved (\S \ref{correlation}).

The reader might wonder whether there are other lensing corrections to
the observed galaxy overdensity. Indeed there are, and they are discussed
further in Appendix \ref{app:lens}. It suffices to say magnification bias
is the dominant effect for regimes of practical interest.

In all illustrative examples below, we employ the following
cosmological parameters: 
the Hubble constant $h = 0.7$, matter density 
$\Omega_m = 0.27$, cosmological constant $\Omega_\Lambda = 0.73$, 
baryon density $\Omega_b = 0.046$, 
power spectrum slope $n = 0.95$ and normalization $\sigma_8 = 0.8$.
We employ the transfer function of \cite{EH98}, and 
the prescription of \cite{smith} for the
nonlinear power spectrum.

\subsection{The Anisotropic Correlation Function}
\label{correlation}

The observed two-point correlation function is:
\begin{eqnarray}
\label{obs}
\xi_{\rm obs} (\chi_1, \thetaB_1; \chi_2, \thetaB_2) =
\langle \delta_{\rm obs} (\chi_1, \thetaB_1) \delta_{\rm obs} (\chi_2, \thetaB_2) \rangle \\ \nonumber
= \xi_{gg} (\chi_1, \thetaB_1; \chi_2, \thetaB_2) + \xi_{g\mu} (\chi_1, \thetaB_1; \chi_2, \thetaB_2) \\ \nonumber
+ \xi_{g\mu} (\chi_2, \thetaB_2; \chi_1, \thetaB_1) + \xi_{\mu\mu} (\chi_1, \thetaB_1; \chi_2, \thetaB_2)
\end{eqnarray}
where
\begin{eqnarray}
\xi_{gg} (1;2) = \langle \delta_g (1) \delta_g (2)\rangle \quad , \quad 
\xi_{g\mu} (1;2) = \langle \delta_g (1) \delta_\mu (2) \rangle \\ \nonumber
\xi_{g\mu} (2;1) = \langle \delta_g (2) \delta_\mu (1) \rangle \quad , \quad
\xi_{\mu\mu} (1;2) = \langle \delta_\mu (1) \delta_\mu (2) \rangle
\end{eqnarray}
with the arguments $1$ and $2$ as shorthands for $\chi_1, \thetaB_1$ and $\chi_2, \thetaB_2$.

Using the Limber approximation, the galaxy-magnification cross-term(s) can be written as:
\begin{eqnarray}
\label{gmuexact}
\xi_{g\mu} (\chi_1, \thetaB_1; \chi_2, \thetaB_2) =
{3\over 2} H_0^2 \Omega_m (5s - 2) \\ \nonumber (1+z_1)
{(\chi_2 - \chi_1) \chi_1 \over \chi_2} \Theta(\chi_1 < \chi_2) \\ \nonumber
\int {d^2 k_\perp \over (2\pi)^2} P_{gm} (z_1, k_\perp) e^{i {\bf k_\perp} \cdot \chi_1 (\thetaB_1 - \thetaB_2)}
\end{eqnarray}
where $\Theta(\chi_1 < \chi_2)$ is a step function which equals $1$ if $\chi_1 < \chi_2$ and vanishes otherwise,
$H_0$ is the Hubble constant today, $\Omega_m$ is the matter density today (normalized by the critical density),
$z_1$ is the redshift corresponding to the comoving distance $\chi_1$, and $P_{gm}$
is the (3D) 
galaxy-mass power spectrum, and ${\bf k_\perp}$ is the transverse Fourier wave vector.
We have used the Poisson equation to relate the gravitational potential $\phi$ to the mass overdensity $\delta$:
\begin{equation}
\nabla^2 \phi = 3 H_0^2 \Omega_m (1+z) \delta/2
\end{equation}
Note that the speed of light is set to $1$ throughout.

The magnification-magnification correlation, or
magnification auto-correlation, is
\begin{eqnarray}
\label{mumuexact}
\xi_{\mu\mu} (\chi_1, \thetaB_1; \chi_2, \thetaB_2) =
[{3\over 2} H_0^2 \Omega_m (5s - 2)]^2 \\ \nonumber
\int_0^{\rm min.(\chi_1,\chi_2)} d\chi' {(\chi_1 - \chi')\chi' \over \chi_1}
{(\chi_2 - \chi')\chi' \over \chi_2} \\ \nonumber
(1+z')^2 \int {d^2 k_\perp \over (2\pi)^2} P_{mm} (z', k_\perp) e^{i {\bf k_\perp} \cdot \chi' (\thetaB_1 - \thetaB_2)}
\end{eqnarray}
where $P_{mm}$ is the (3D) mass-mass power spectrum: 
it is evaluated at the redshift $z'$ which
corresponds to the integration variable $\chi'$.

The focus in the literature has been on $\xi_{\rm obs}$ as an angular correlation:
$\chi_1$ and $\chi_2$ are typically integrated over some radial selection functions, and
they can signify either two different redshift bins (i.e. angular cross-correlation between
galaxies/quasars at two different redshifts \cite{MJ98}), or the same redshift bin (i.e.
angular auto-correlation function \cite{MJV98}).

Here, let us take a slightly different perspective: think of $\xi_{\rm obs}$ as a 3D correlation
function \cite{matsubara00}. For a galaxy survey with redshift information (either spectroscopic
redshifts or high quality photometric redshifts), this would be a very natural thing to do.
Further, suppose one has a galaxy survey, or a subsample thereof, that 
spans some finite redshift range such that the radial separation
$\chi_1 - \chi_2$ is always small compared to 
$\chi_1$ or $\chi_2$. This is a sensible assumption since at sufficiently large
separations, galaxy evolution becomes important and complicates one's analysis.
Let $\bar\chi$ be the mean radial comoving distance to these galaxies, and $\bar z$ be
the associated mean redshift.
The galaxy-magnification cross-correlation and the magnification auto-correlation can
be simplified as follows:
\begin{eqnarray}
\label{gmu}
\xi_{g\mu} (\chi_1, \thetaB_1; \chi_2, \thetaB_2) + \xi_{g\mu} (\chi_2, \thetaB_2; \chi_1, \thetaB_1) =
\\ \nonumber 
{3\over 2} H_0^2 \Omega_m (5s - 2) (1+\bar z)
| \chi_2 - \chi_1 | \\ \nonumber
\int {d^2 k_\perp \over (2\pi)^2} P_{gm} (\bar z, k_\perp) 
e^{i {\bf k_\perp} \cdot \bar\chi (\thetaB_1 - \thetaB_2)}
\end{eqnarray}
\begin{eqnarray}
\label{mumu}
\xi_{\mu\mu} (\chi_1, \thetaB_1; \chi_2, \thetaB_2) =
[{3\over 2} H_0^2 \Omega_m (5s - 2)]^2 \\ \nonumber
\int_0^{\bar\chi} d\chi' \left[{(\bar\chi - \chi')\chi' \over \bar\chi}\right]^2 (1+z')^2 \\ \nonumber
\int {d^2 k_\perp \over (2\pi)^2} P_{mm} (z', k_\perp) e^{i {\bf k_\perp} \cdot \chi' (\thetaB_1 - \thetaB_2)}
\end{eqnarray}
where we have Taylor expanded $\chi_1$ and $\chi_2$ around $\bar\chi$ and retained the lowest
order contributions. It is useful to compare these two expressions to the
intrinsic (unlensed) galaxy auto-correlation, or galaxy-galaxy correlation:
\begin{eqnarray}
\label{gg}
&& \xi_{gg} (\chi_1, \thetaB_1; \chi_2, \thetaB_2) = \\ \nonumber &&
\xi_{gg} (\sqrt{ (\chi_1-\chi_2)^2 + 
\bar\chi^2 (\thetaB_1 - \thetaB_2)^2}) = \\ \nonumber &&
\int {d^3 k \over (2\pi)^3} P_{gg} (\bar z, k) e^{i {\bf k} \cdot ({\bf x_1} - {\bf x_2})}
\end{eqnarray}
ignoring for now the issue of redshift distortion, 
which will be addressed in Paper II. Note that ${\bf x_1}$ and ${\bf x_2}$ refer
to the points corresponding to $\chi_1, \thetaB_1$ and $\chi_2, \thetaB_2$.

The observed correlation function is a sum of all three correlations above (eq. [\ref{obs}], 
[\ref{gmu}], [\ref{mumu}] and [\ref{gg}]). 
(A discussion of higher order corrections to the latter three 
can be found in Appendix \ref{app:lens2}.)
Viewed in this way, the anisotropy of the lensing induced corrections
is quite striking: $\xi_{g\mu}(1,2) + \xi_{g\mu}(2,1)$ scales linearly with the line-of-sight (LOS) separation
$|\chi_2 - \chi_1|$
(i.e. it increases rather than decreases with the separation!), and 
$\xi_{\mu\mu}$ is independent of the LOS separation.
The intrinsic galaxy auto-correlation
$\xi_{gg}$ is isotropic and generally decreases with separation.

\begin{figure}[tb]
\centerline{\epsfxsize=9cm\epsffile{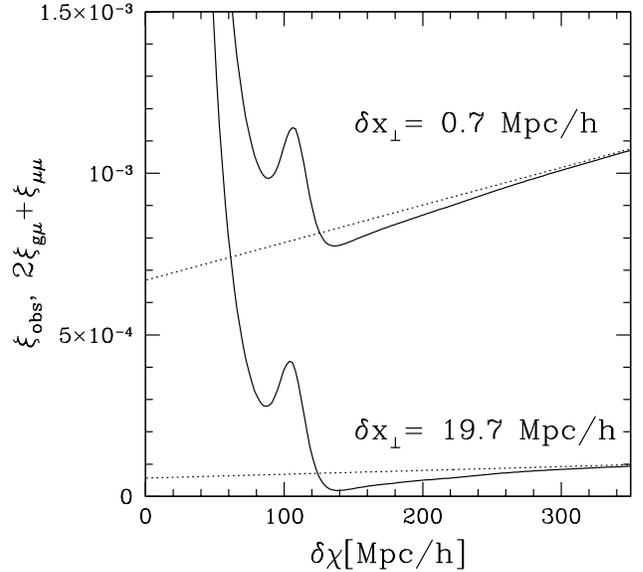}}
\caption{An illustration of how $\xi_{gg}, \xi_{g\mu}$ and $\xi_{\mu\mu}$ in principle can be
obtained from the observed $\xi_{\rm obs}$. At a given transverse separation $\delta x_\perp$
(2 examples are shown), $\xi_{\rm obs}$ (solid line) at a large LOS separation $\delta\chi$
is dominated by the magnification corrections $2\xi_{g\mu} + \xi_{\mu\mu}$ which have
the form $f(\delta x_\perp) \delta\chi + g(\delta x_\perp)$ i.e. it is linear in
$\delta\chi$ (dotted line). The extrapolation of this dotted line to $\delta\chi=0$
gives $g$ or equivalently $\xi_{\mu\mu}$. Its slope gives
$f$ which can be multiplied by $\delta\chi$ to obtain $2\xi_{g\mu}$. 
Finally, subtracting the dotted
line from $\xi_{\rm obs}$ yields $\xi_{gg}$. 
}
\label{xi.fit}
\end{figure}

\begin{figure*}[tb]
\subfigure[
]
{
	\label{xicontour.new.0.35}
	\includegraphics[width=.41\textwidth]{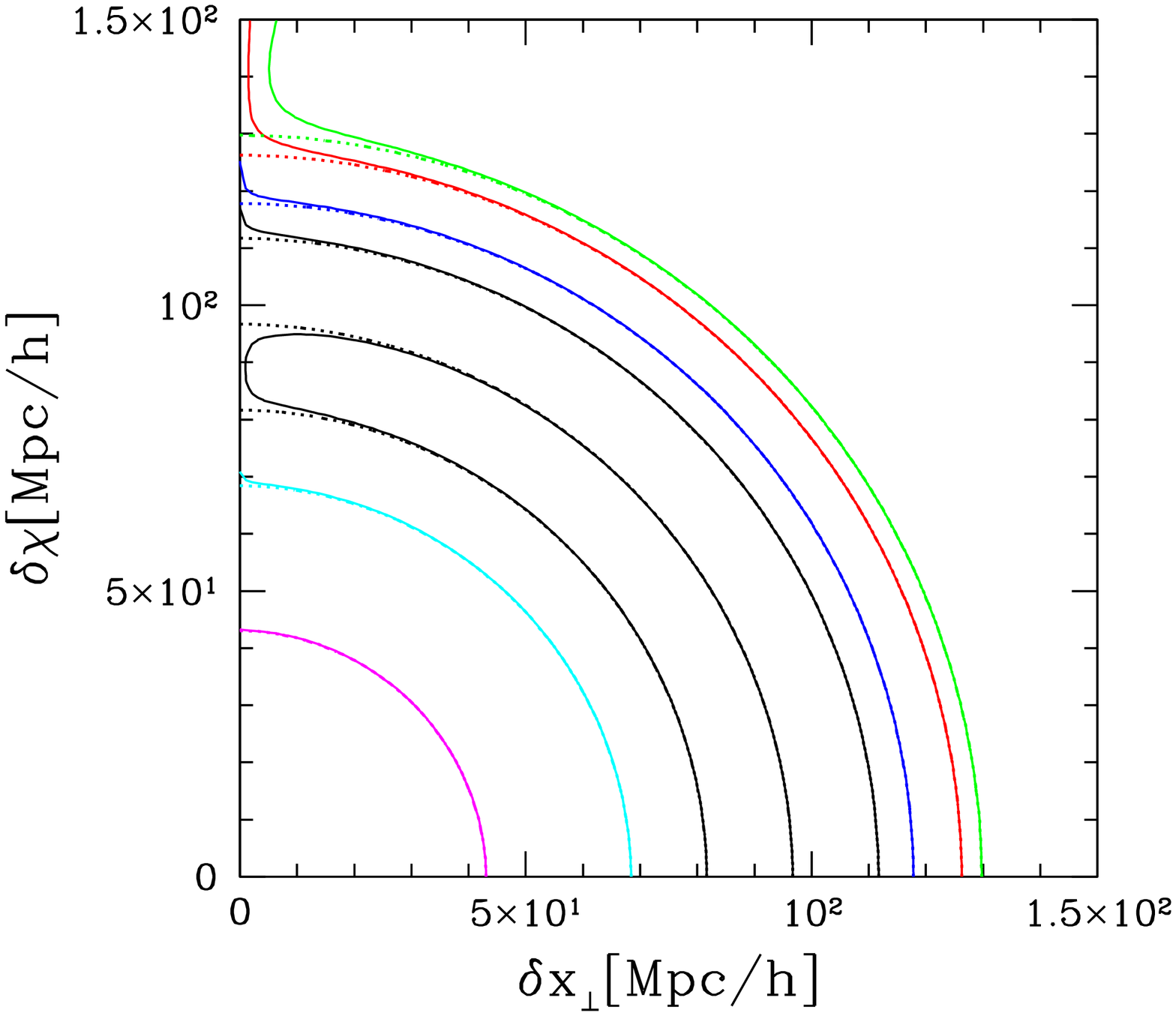}
}
\hspace{0.01in}
\subfigure[
]
{
	\label{xi.try6.0.35}
	\includegraphics[width=.41\textwidth]{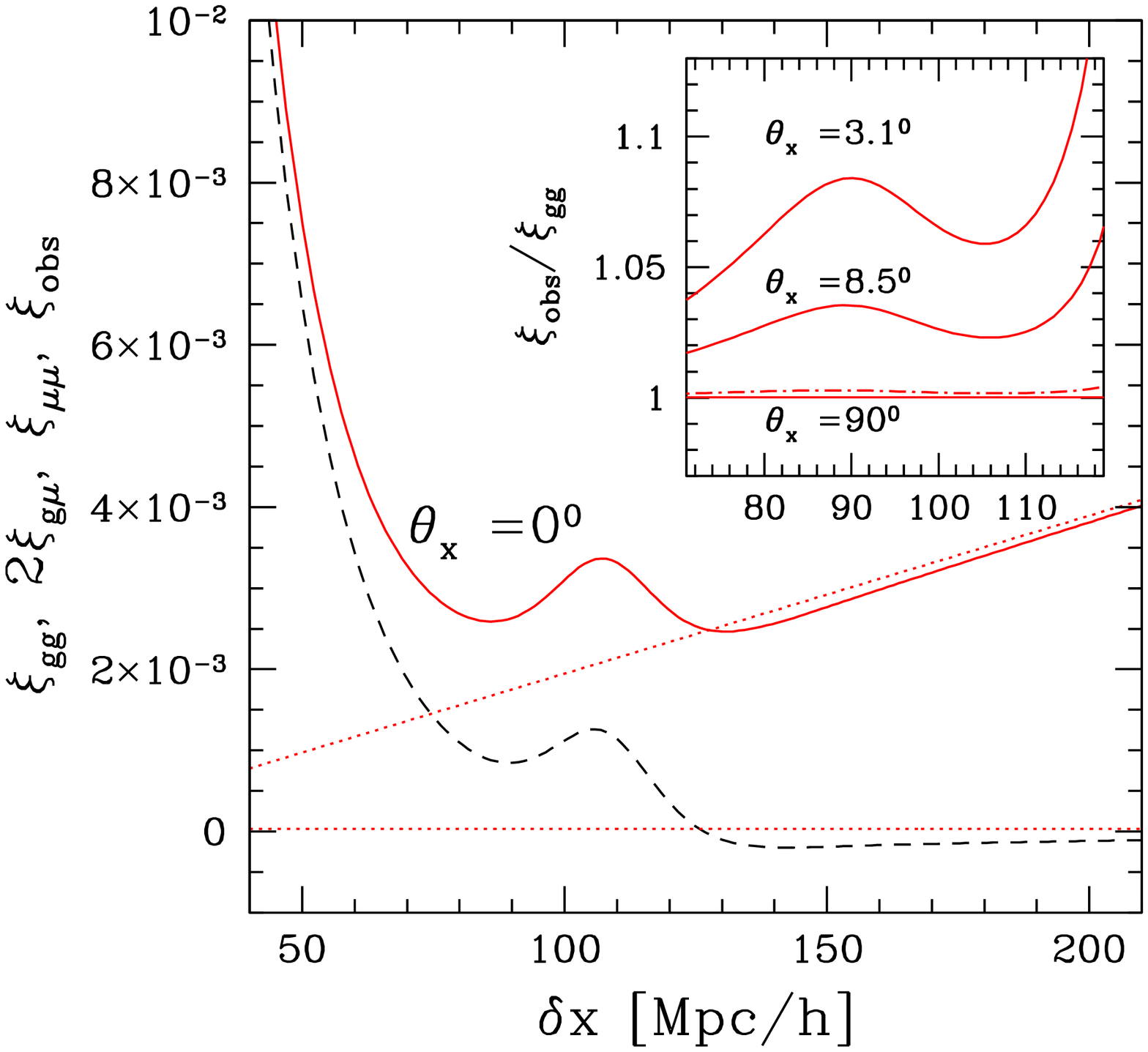}
}
\vspace{0.01in}
\subfigure[
]
{
	\label{xipeakcheck.curv.combo.0.35}
	\includegraphics[width=.41\textwidth]{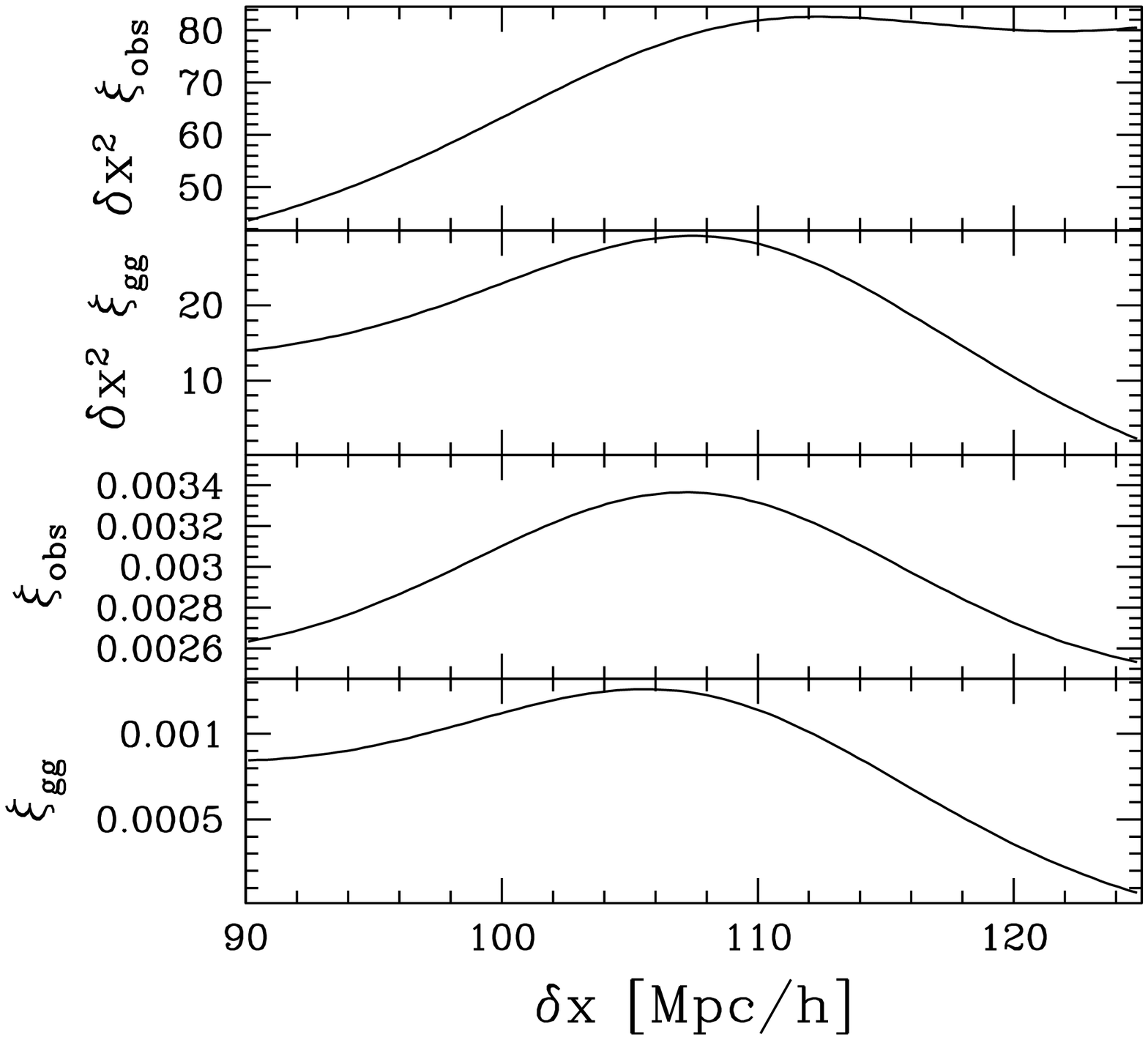}
}
\hspace{0.01in}
\subfigure[
]
{
	\label{xi.try6.mono.0.35}
	\includegraphics[width=.41\textwidth]{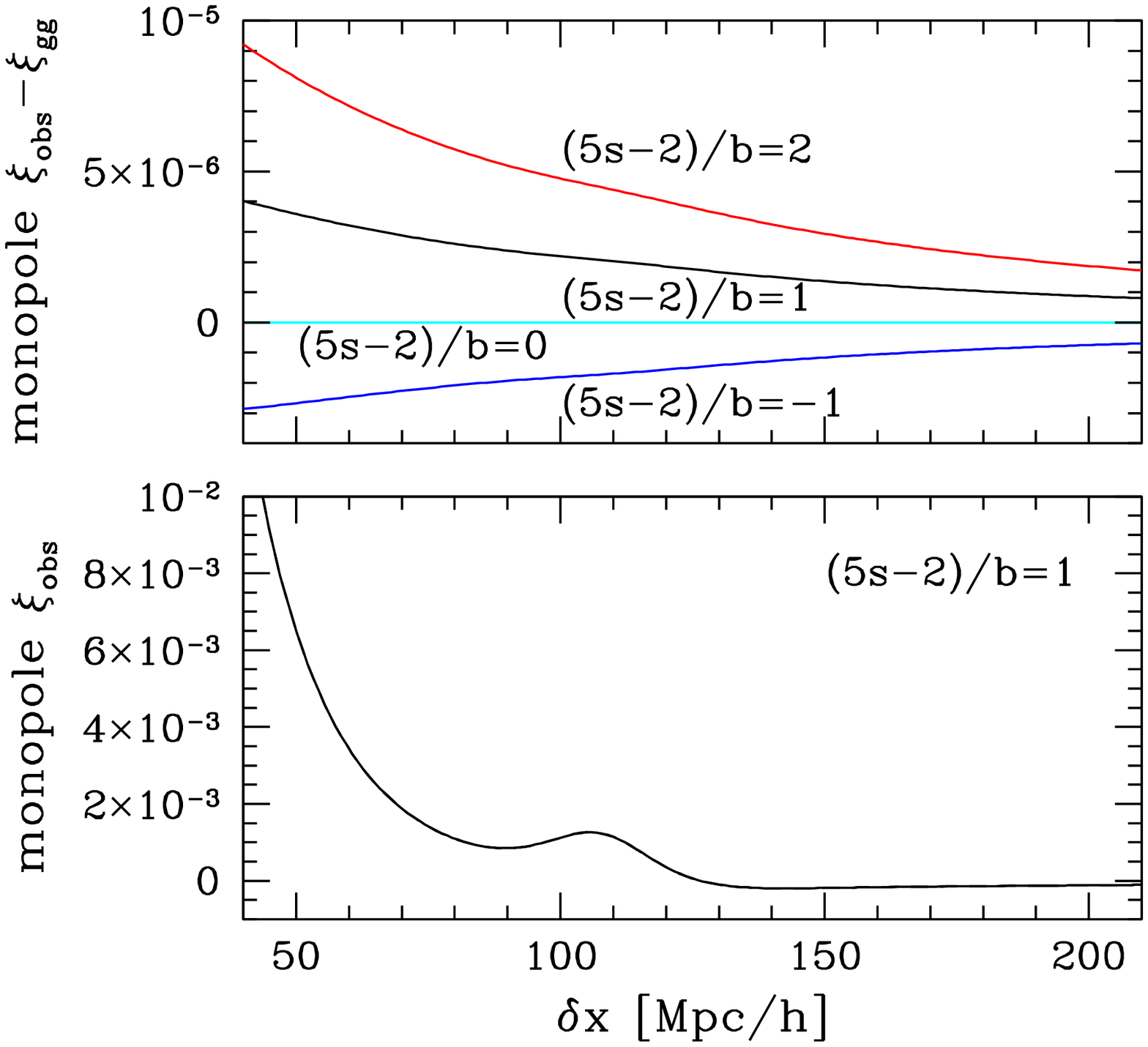}
}
\caption{$\bar z=0.35$: ({\bf a}) Contours of constant $\xi_{\rm obs}$ (solid) and
$\xi_{gg}$ (dotted), left to right: 
0.01 (magenta), 0.002 (cyan), 0.001 (black; triple contours), 0.0005 (blue), 0 (red), -0.0001 (green).
The LOS separation is $\delta\chi$; the transverse separation is $\delta x_\perp$. 
$\xi_{gg}$ is isotropic but $\xi_{\rm obs}$ is not. 
({\bf b}) $\xi_{gg}$ (black dashed), $2 \xi_{g\mu}$ (sloped red dotted), $\xi_{\mu\mu}$ (flat
red dotted) and $\xi_{\rm obs}$ (red solid) for a separation vector oriented along the LOS. The inset
shows the ratio $\xi_{\rm obs}/\xi_{gg}$ for several other orientations (solid);
dot-dashed line shows the ratio of the respective monopoles. 
Note $\delta x^2 = \delta \chi^2 + \delta x_\perp^2$.
({\bf c}) A zoomed in view of $\xi_{gg}, \xi_{\rm obs}, \delta x^2 \xi_{gg}, \delta x^2 \xi_{\rm obs}$ 
around the baryon wiggle for a separation vector oriented along the LOS.
Note how dangerous it is to use $\delta x^2 \xi_{\rm obs}$ to locate the baryon peak.
({\bf d}) Lower panel shows the monopole of $\xi_{\rm obs}$; upper panel shows the 
difference monopole $\xi_{\rm obs} - \xi_{gg}$ for several different 
values of $(5s-2)/b$. 
Unless otherwise stated (as in panel d), $(5s-2)/b = 1$ throughout.
All $\xi$'s are normalized by $b^2$. 
}
\label{xiall0.35}
\end{figure*}

\begin{figure*}[tb]
\subfigure[
]
{
	\label{xicontour.new.1}
	\includegraphics[width=.41\textwidth]{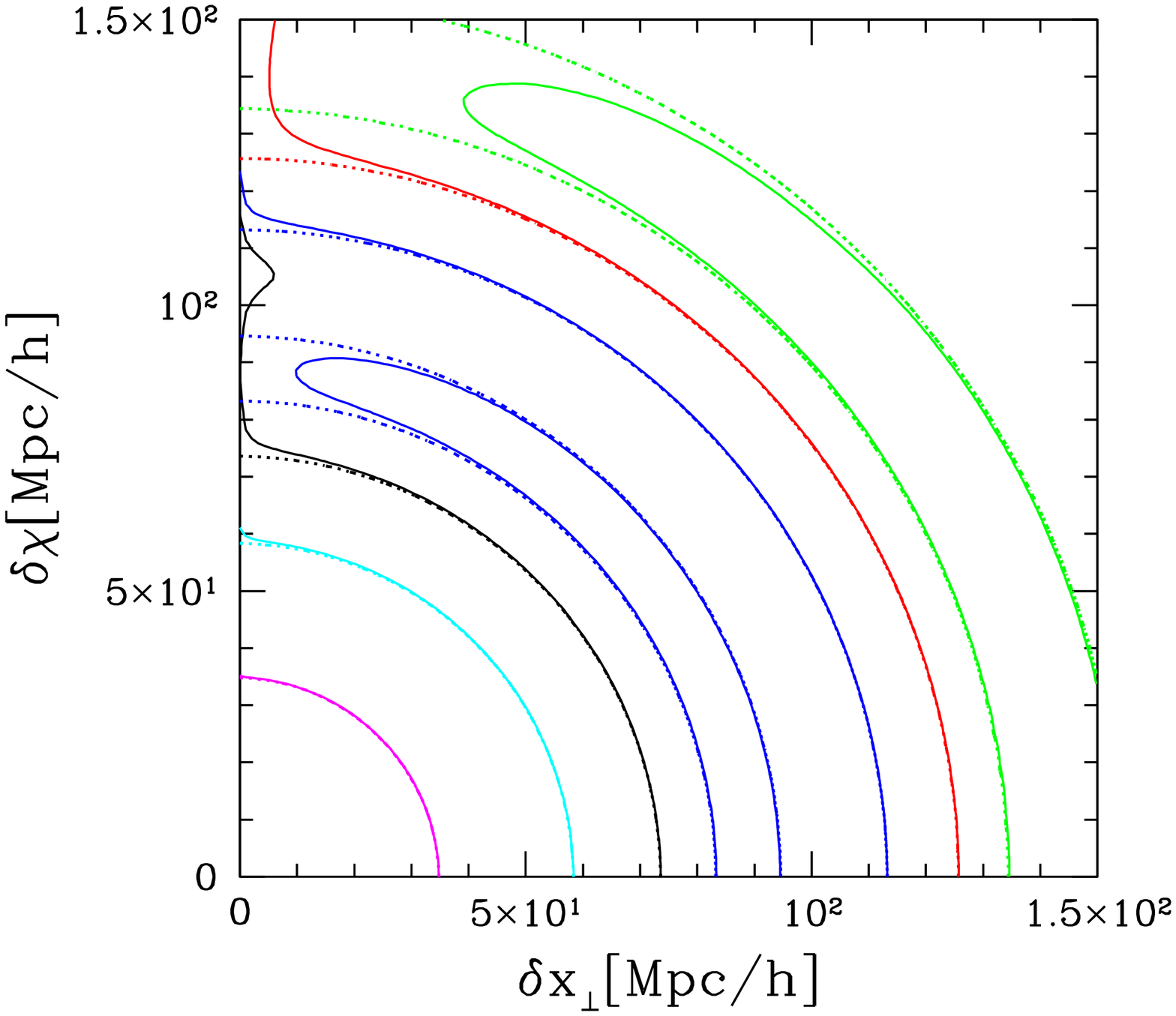}
}
\hspace{0.01in}
\subfigure[
]
{
	\label{xi.try6.1}
	\includegraphics[width=.41\textwidth]{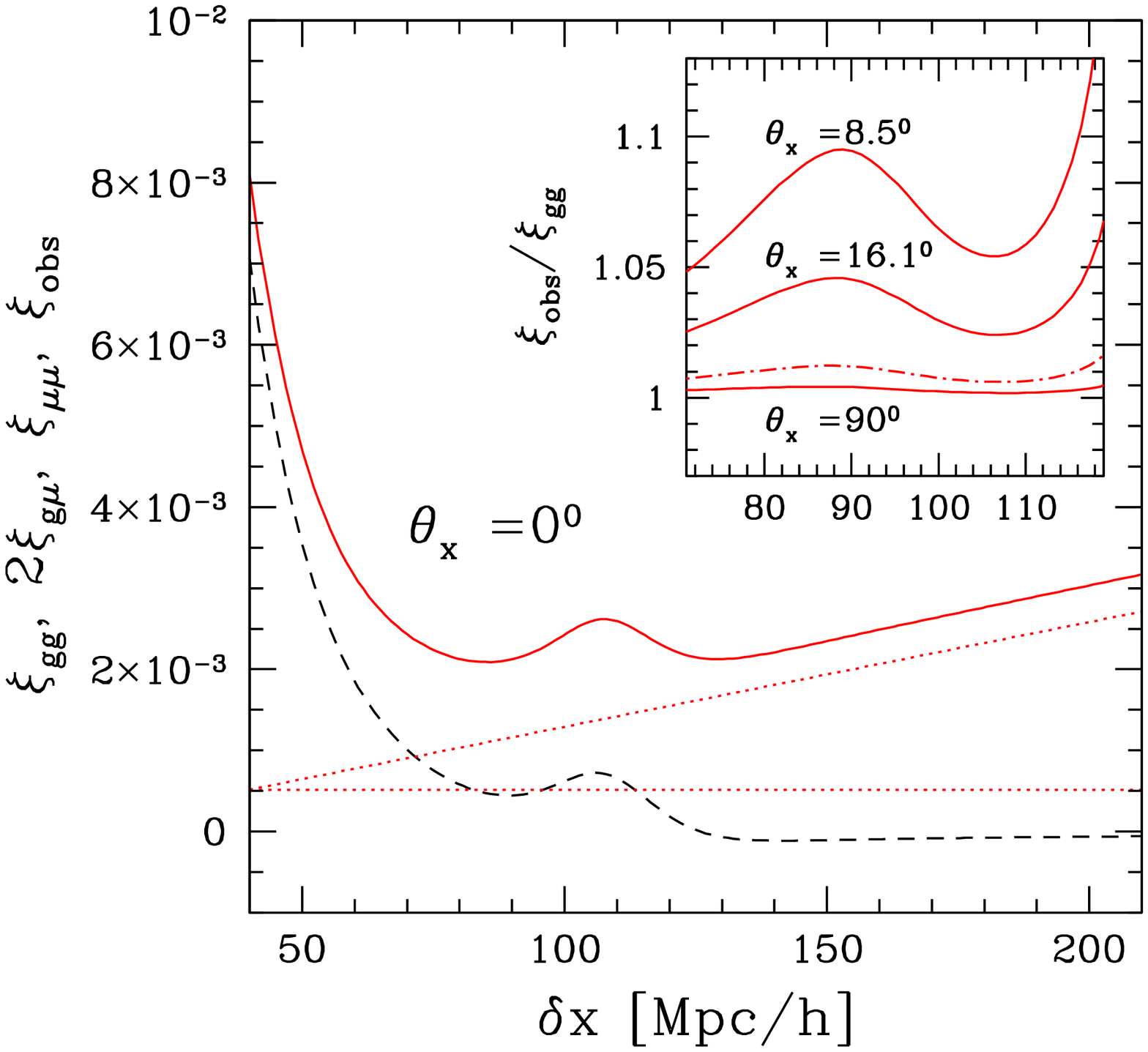}
}
\vspace{0.01in}
\subfigure[
]
{
	\label{xipeakcheck.curv.combo.1}
	\includegraphics[width=.41\textwidth]{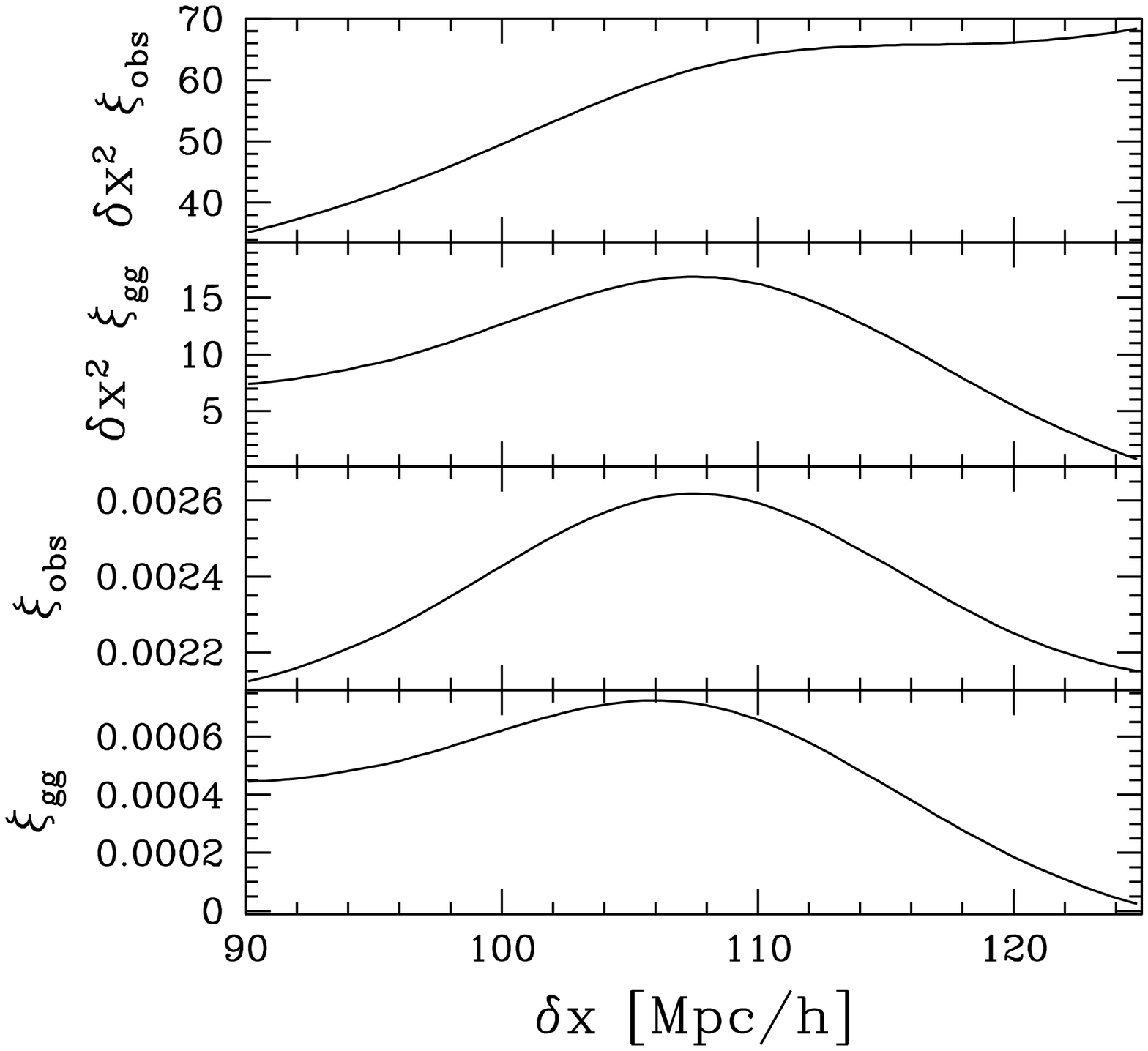}
}
\hspace{0.01in}
\subfigure[
]
{
	\label{xi.try6.mono.1}
	\includegraphics[width=.41\textwidth]{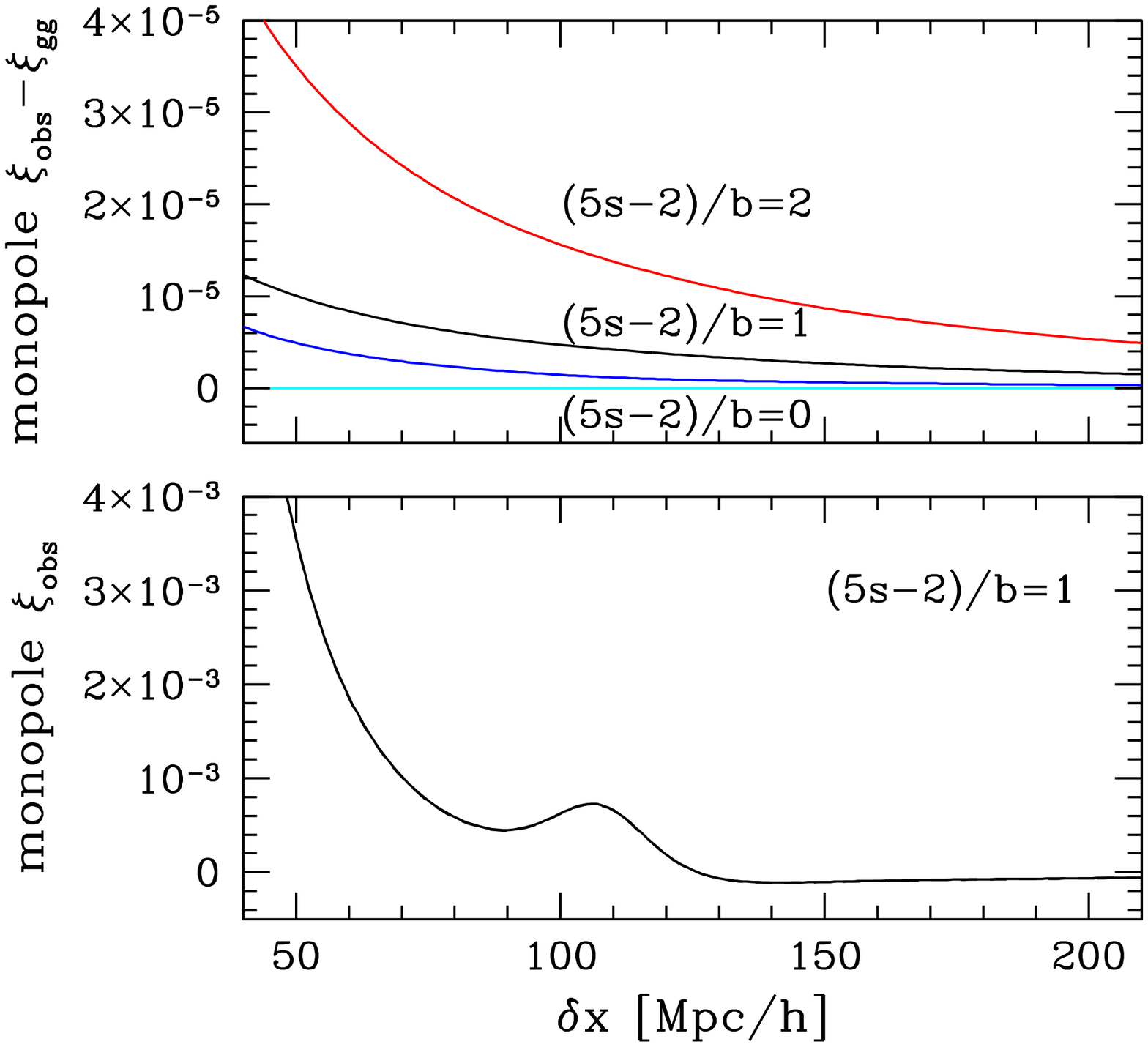}
}
\caption{Analog of Fig. \ref{xiall0.35} for $\bar z=1$.
The contours in (a) are: 0.01 (magenta), 0.002 (cyan), 0.0008 (black),
0.0005 (blue; triple contours), 0 (red) and -0.001 (green; double contours).
}
\label{xiall1}
\end{figure*}

\begin{figure*}[tb]
\subfigure[
]
{
	\label{xicontour.new.1.5}
	\includegraphics[width=.41\textwidth]{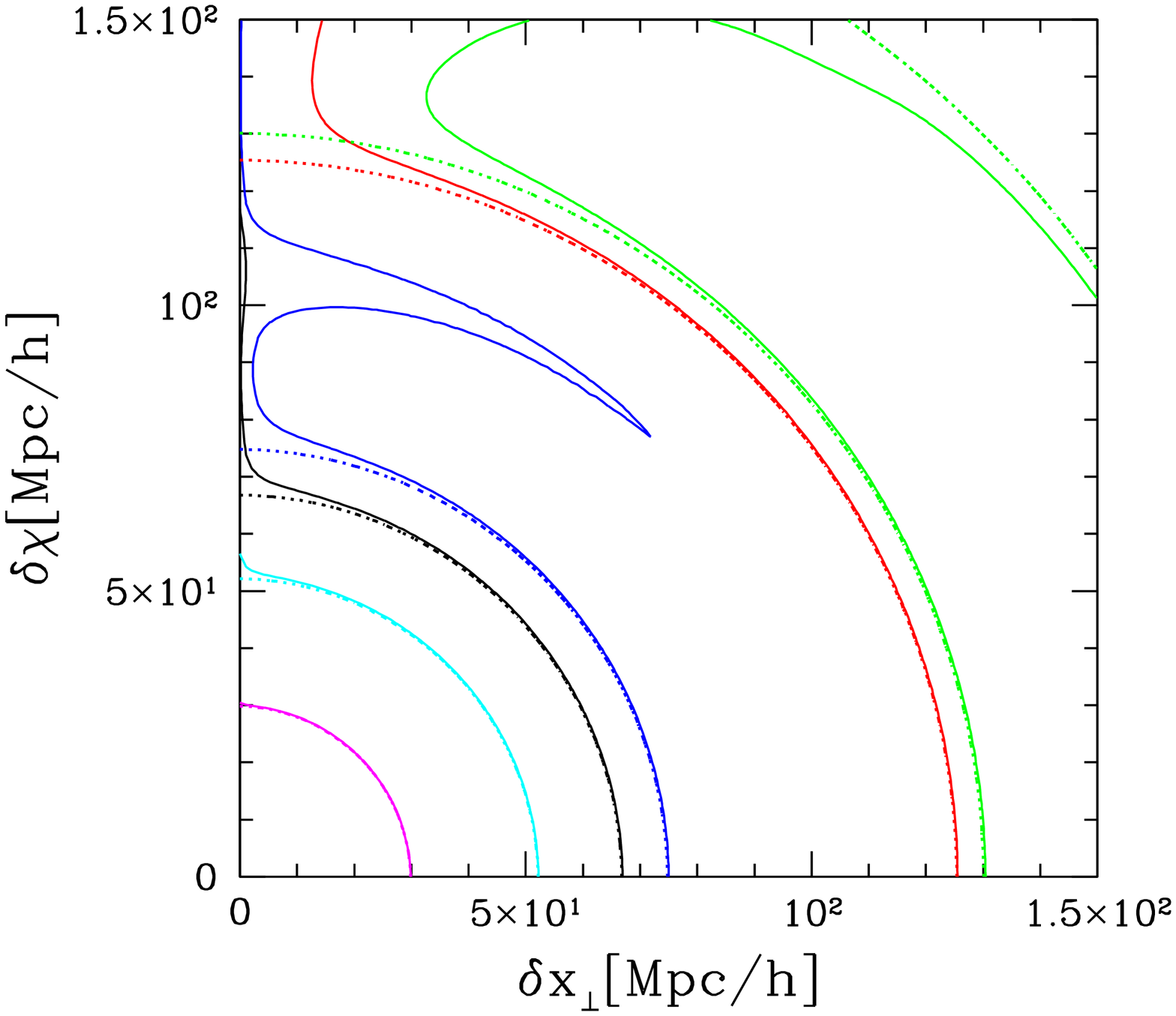}
}
\hspace{0.01in}
\subfigure[
]
{
	\label{xi.try6.1.5}
	\includegraphics[width=.41\textwidth]{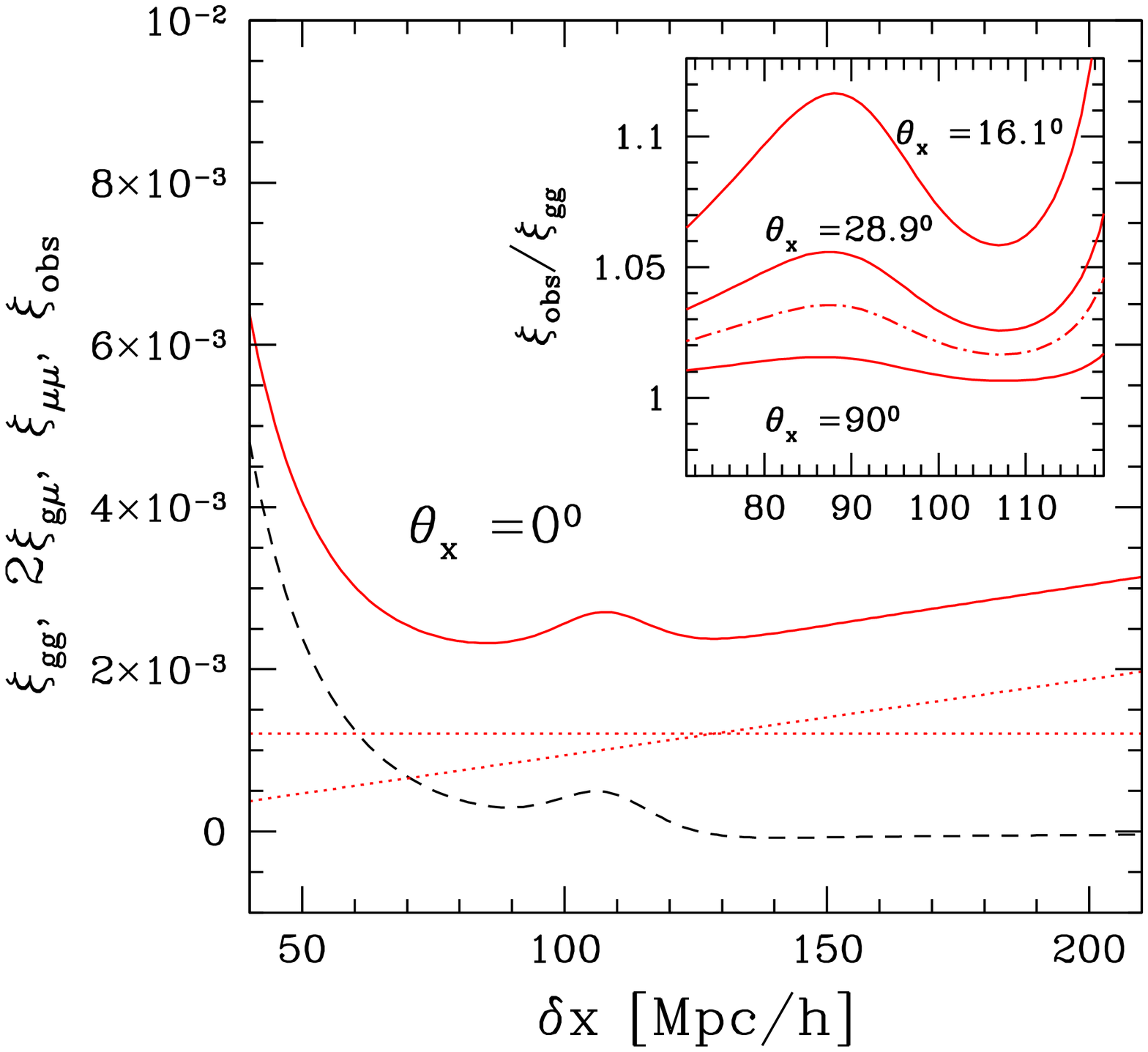}
}
\vspace{0.01in}
\subfigure[
]
{
	\label{xipeakcheck.curv.combo.1.5}
	\includegraphics[width=.41\textwidth]{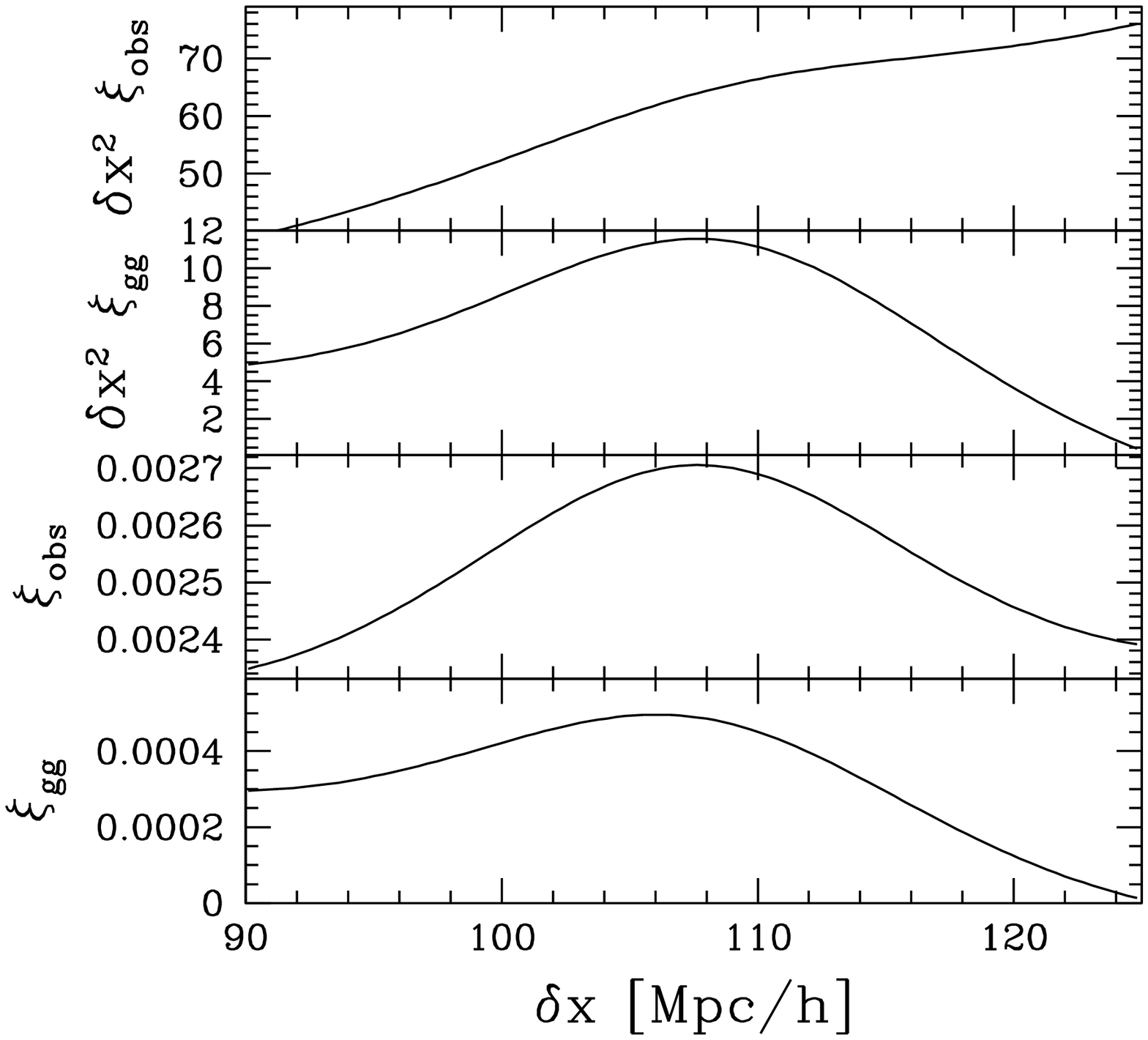}
}
\hspace{0.01in}
\subfigure[
]
{
	\label{xi.try6.mono.1.5}
	\includegraphics[width=.41\textwidth]{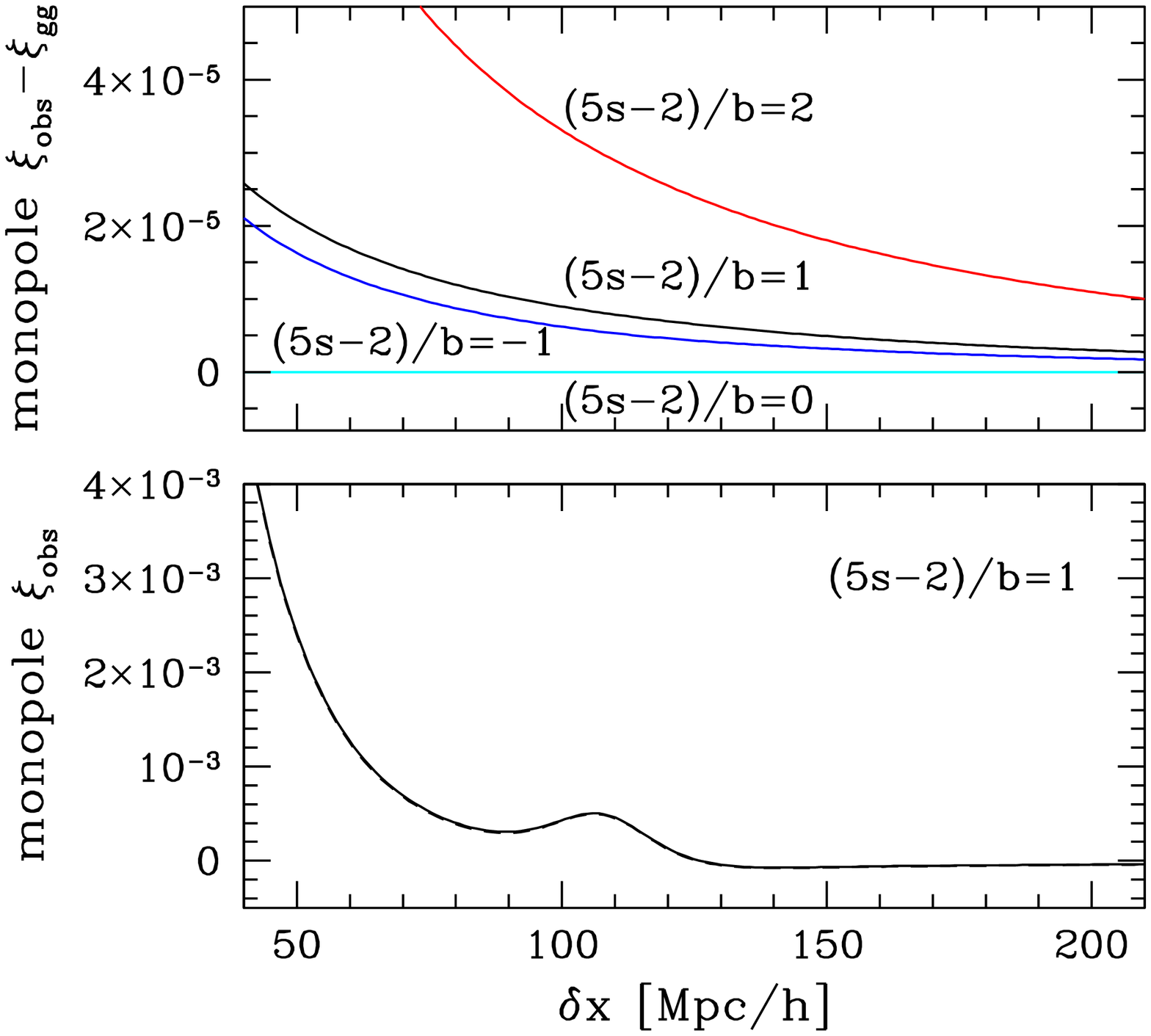}
}
\caption{Analog of Fig. \ref{xiall0.35} for $\bar z=1.5$.
The contours in (a) are: 0.01 (magenta), 0.002 (cyan), 0.0008 (black),
0.0005 (blue), 0 (red) and -0.00005 (green; double contours).
}
\label{xiall1.5}
\end{figure*}

\begin{figure*}[tb]
\subfigure[
]
{
	\label{xicontour.new.2}
	\includegraphics[width=.41\textwidth]{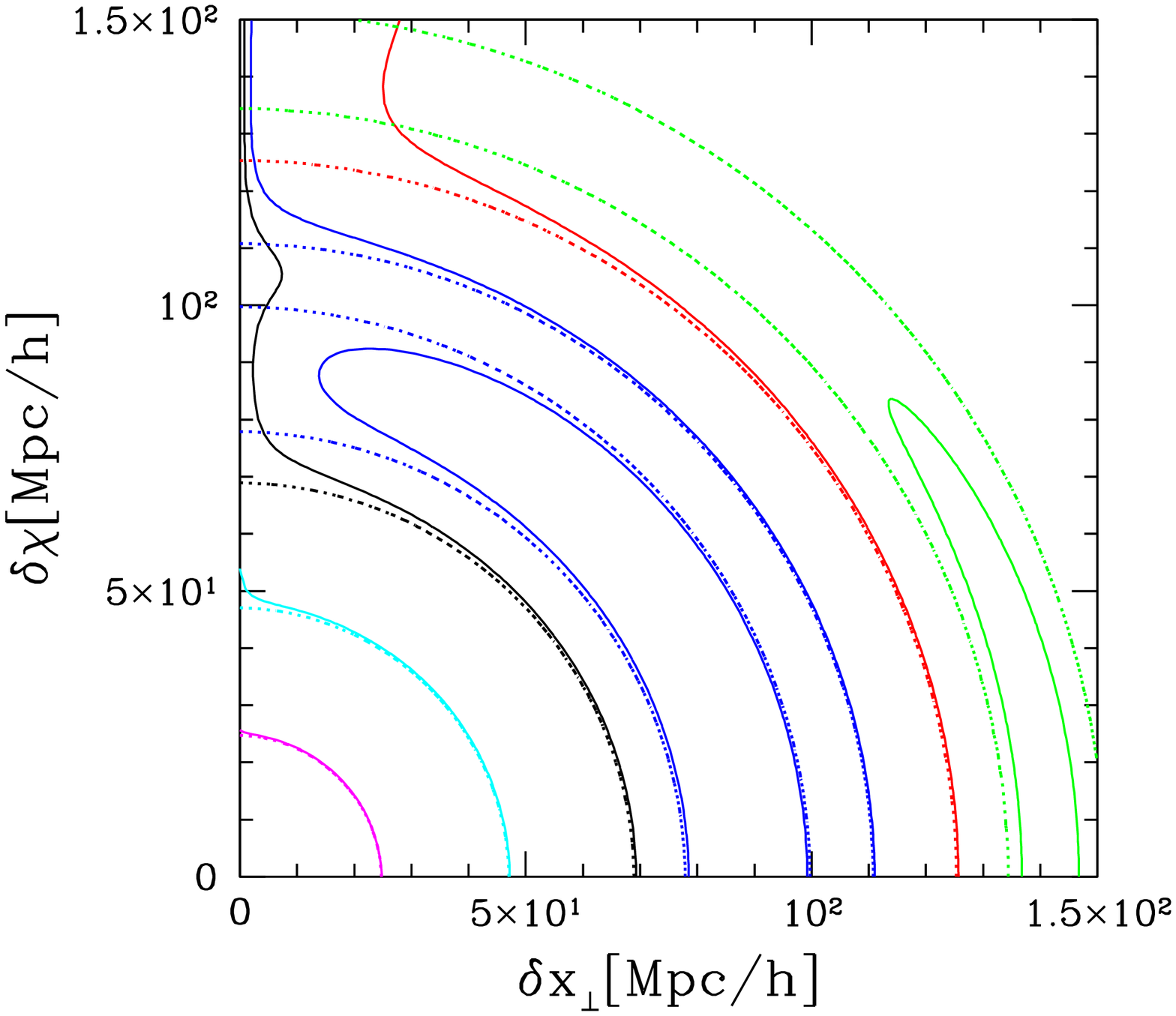}
}
\hspace{0.01in}
\subfigure[
]
{
	\label{xi.try6.2}
	\includegraphics[width=.41\textwidth]{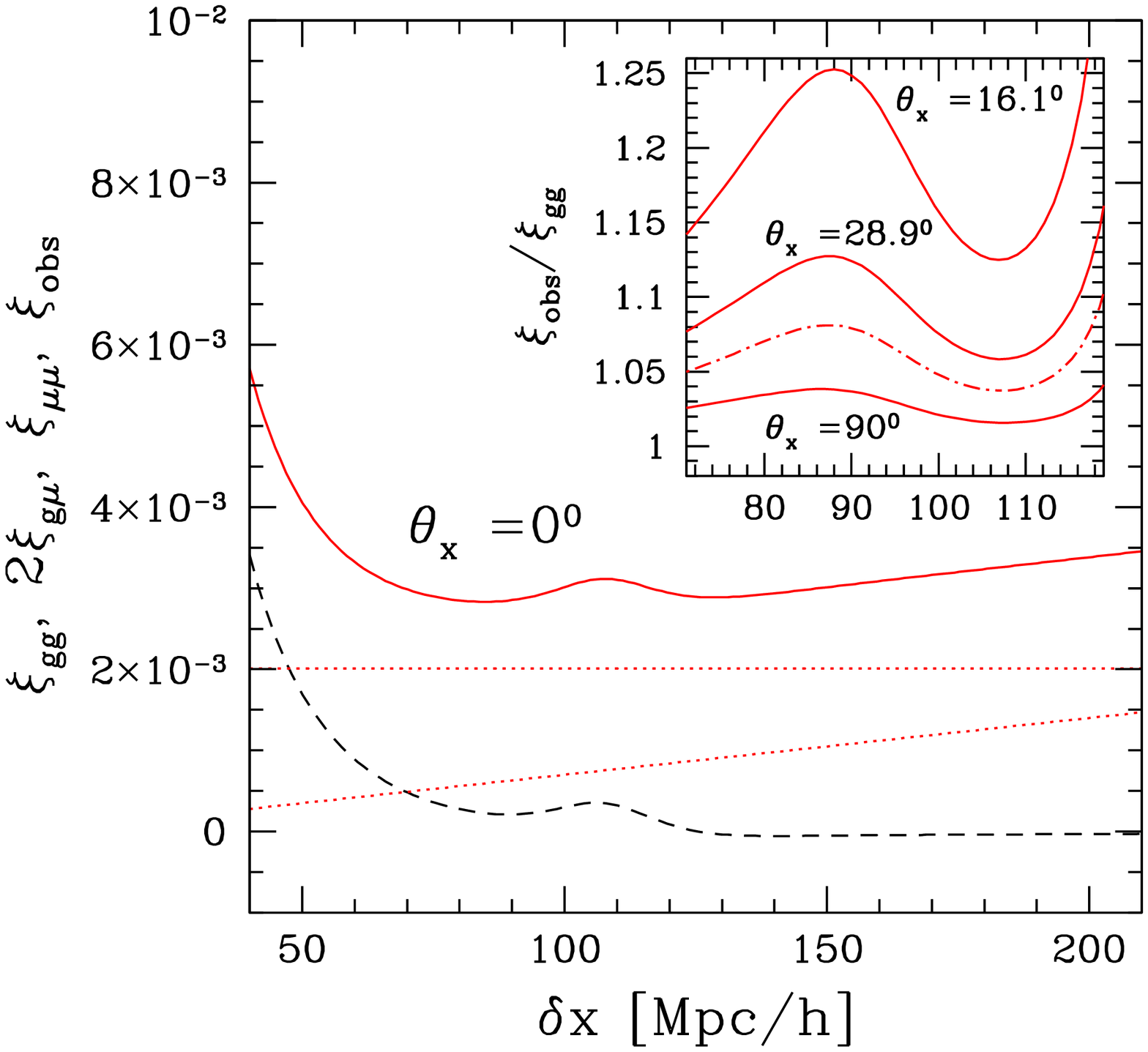}
}
\vspace{0.01in}
\subfigure[
]
{
	\label{xipeakcheck.curv.combo.2}
	\includegraphics[width=.41\textwidth]{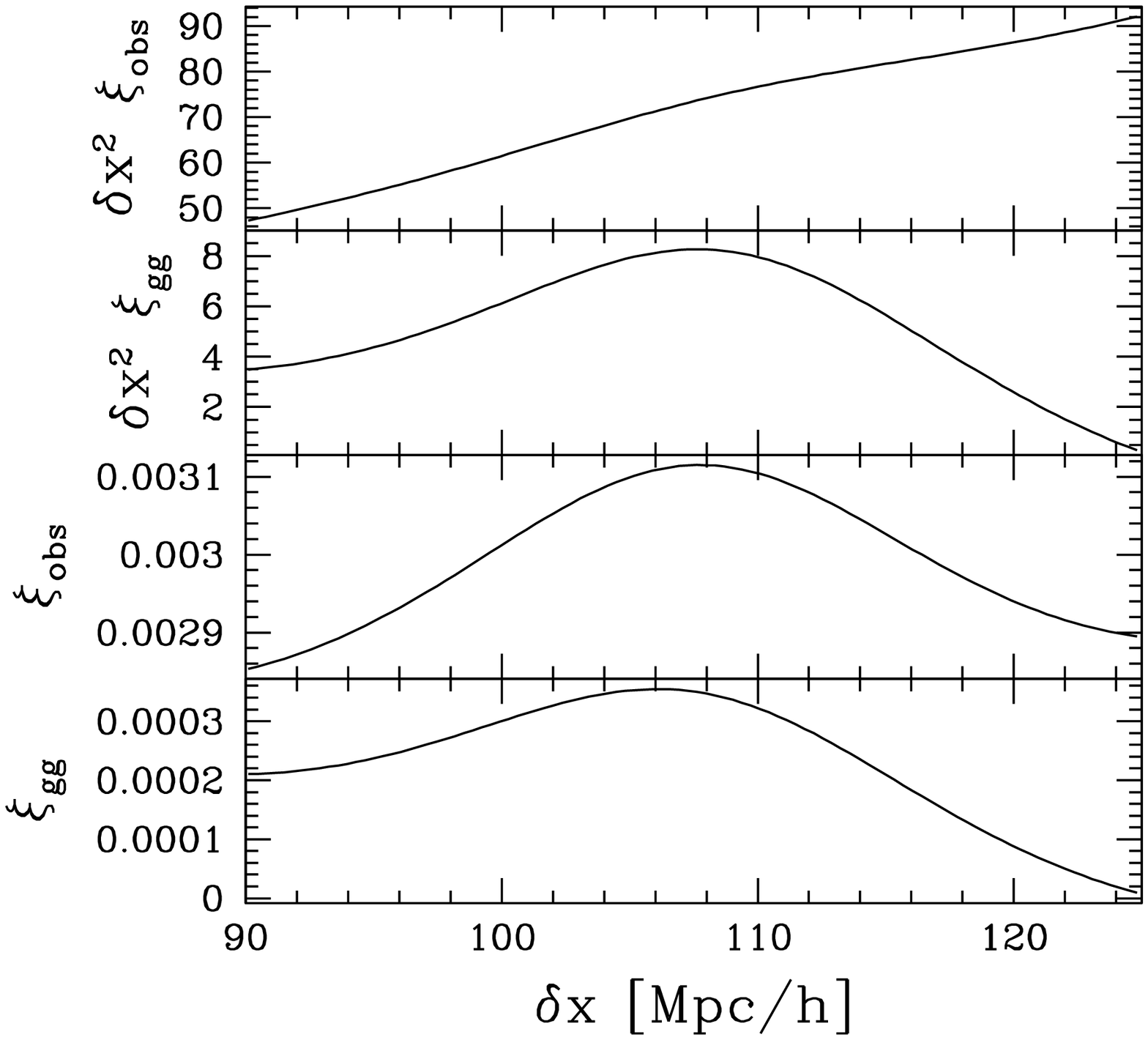}
}
\hspace{0.01in}
\subfigure[
]
{
	\label{xi.try6.mono.2}
	\includegraphics[width=.41\textwidth]{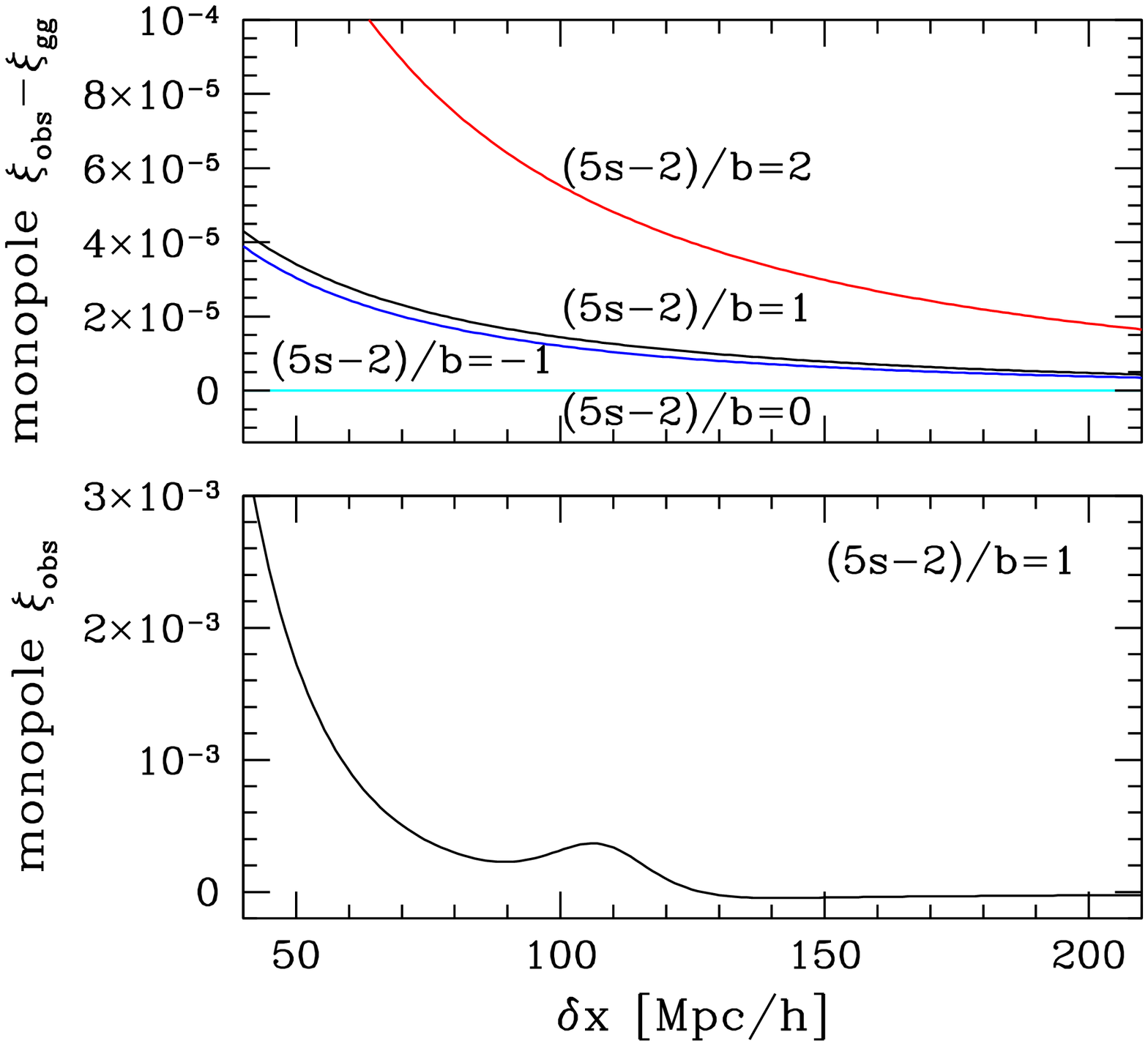}
}
\caption{Analog of Fig. \ref{xiall0.35} for $\bar z=2$.
The contours in (a) are: 0.01 (magenta), 0.002 (cyan), 0.0005 (black),
0.0003 (blue; triple contours), 0 (red) and -0.00005 (green; double contours).
}
\label{xiall2}
\end{figure*}

\begin{figure*}[tb]
\subfigure[
]
{
	\label{xicontour.new.3}
	\includegraphics[width=.41\textwidth]{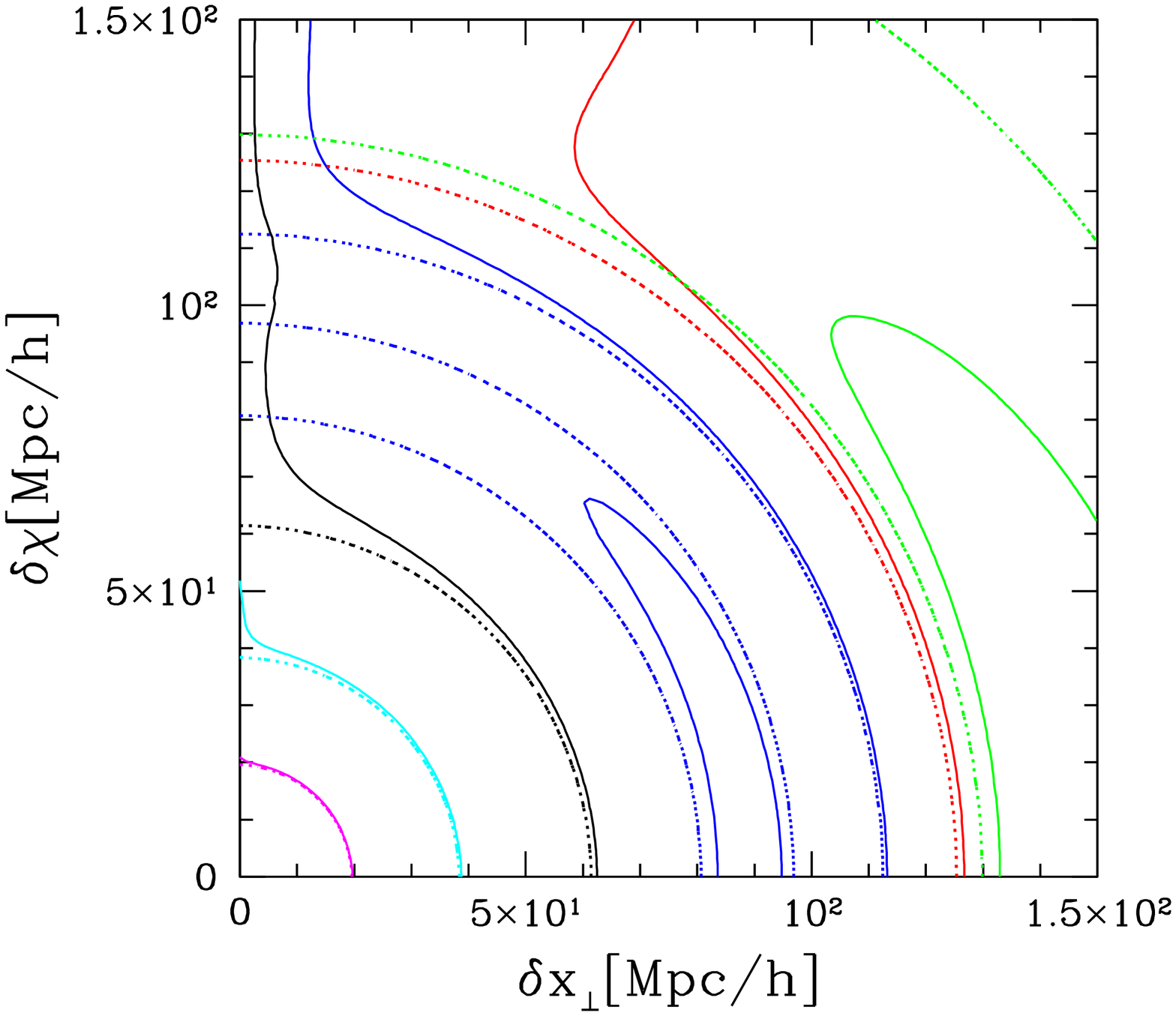}
}
\hspace{0.01in}
\subfigure[
]
{
	\label{xi.try6.3}
	\includegraphics[width=.41\textwidth]{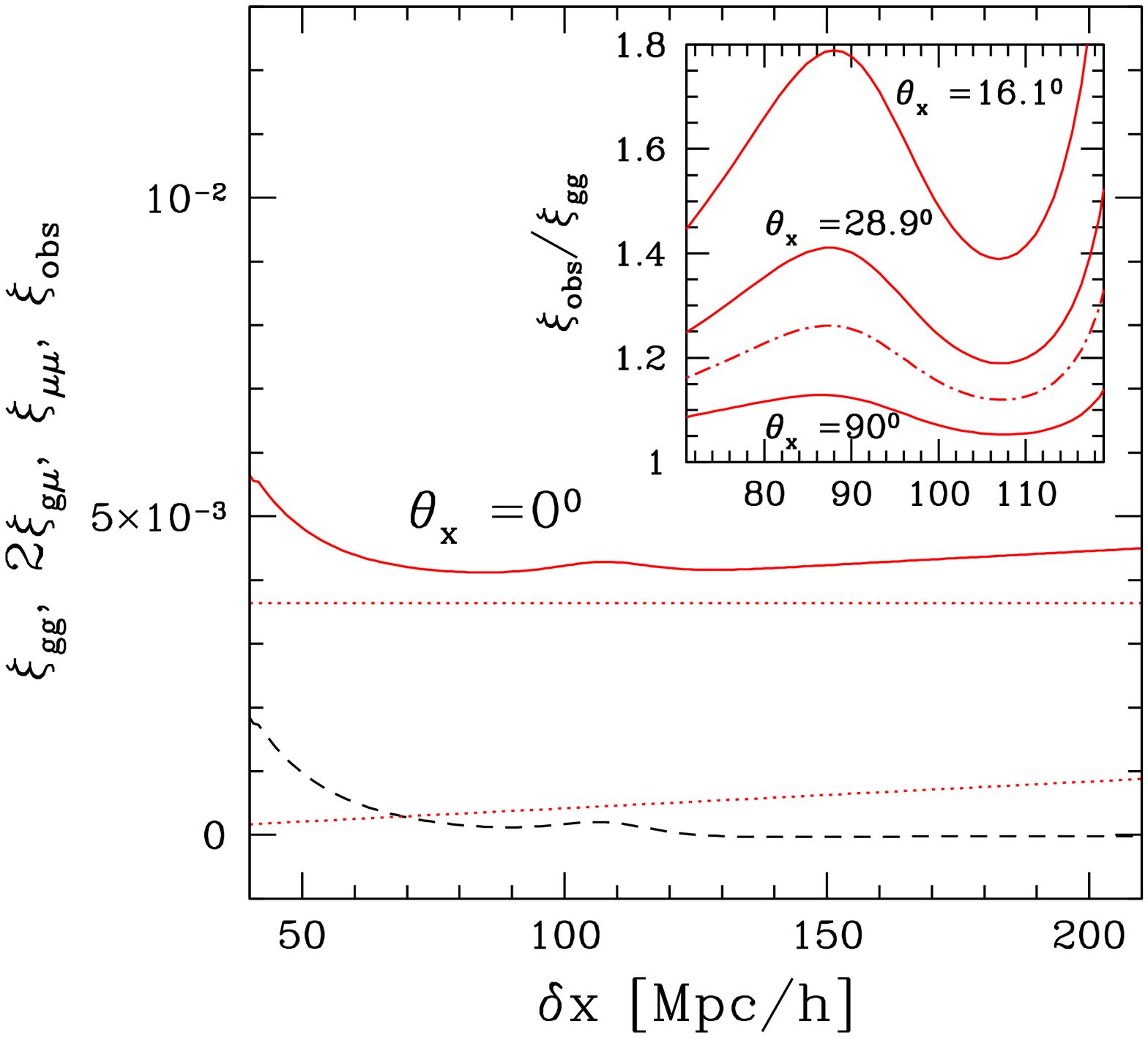}
}
\vspace{0.01in}
\subfigure[
]
{
	\label{xipeakcheck.curv.combo.3}
	\includegraphics[width=.41\textwidth]{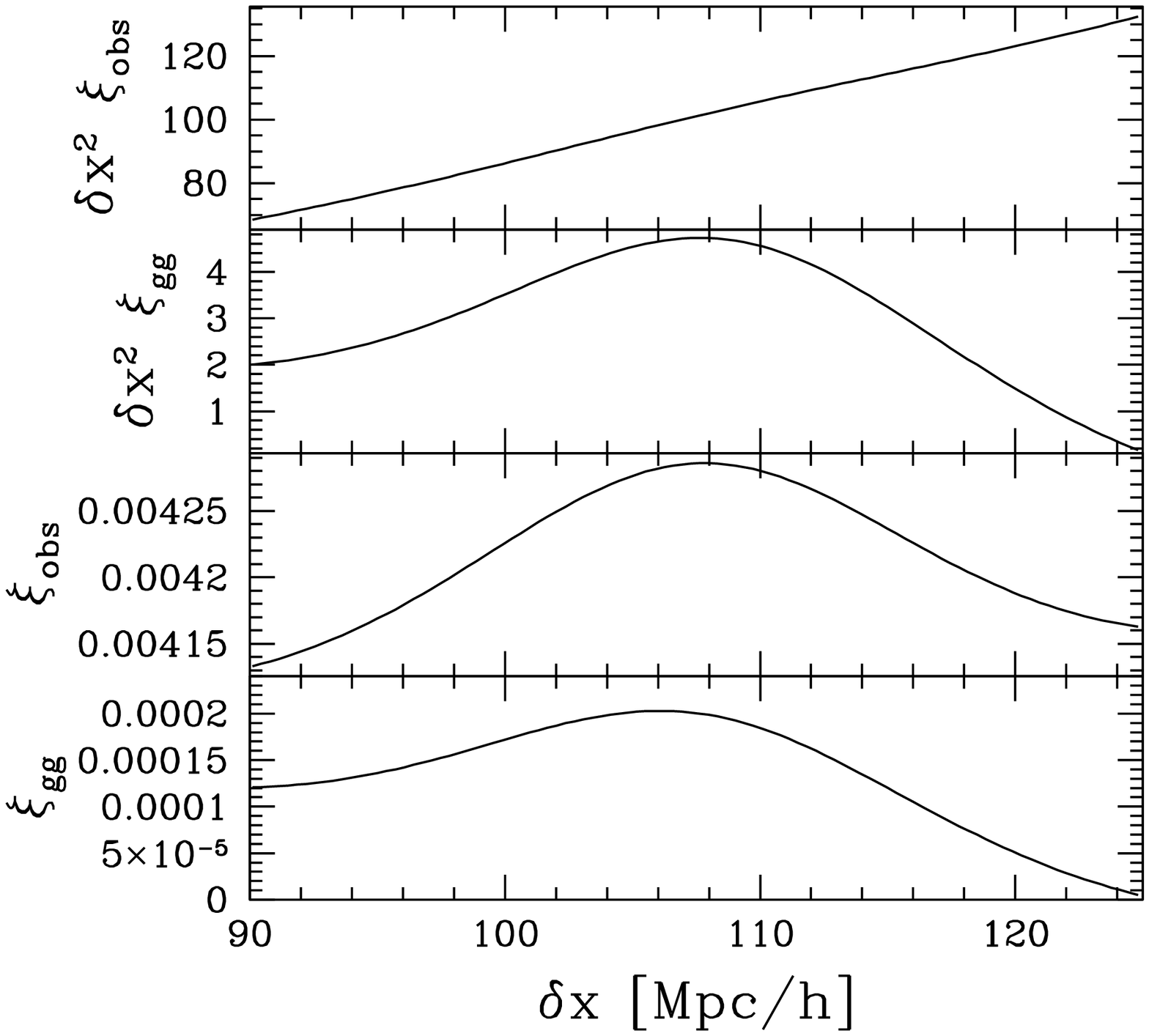}
}
\hspace{0.01in}
\subfigure[
]
{
	\label{xi.try6.mono.3}
	\includegraphics[width=.41\textwidth]{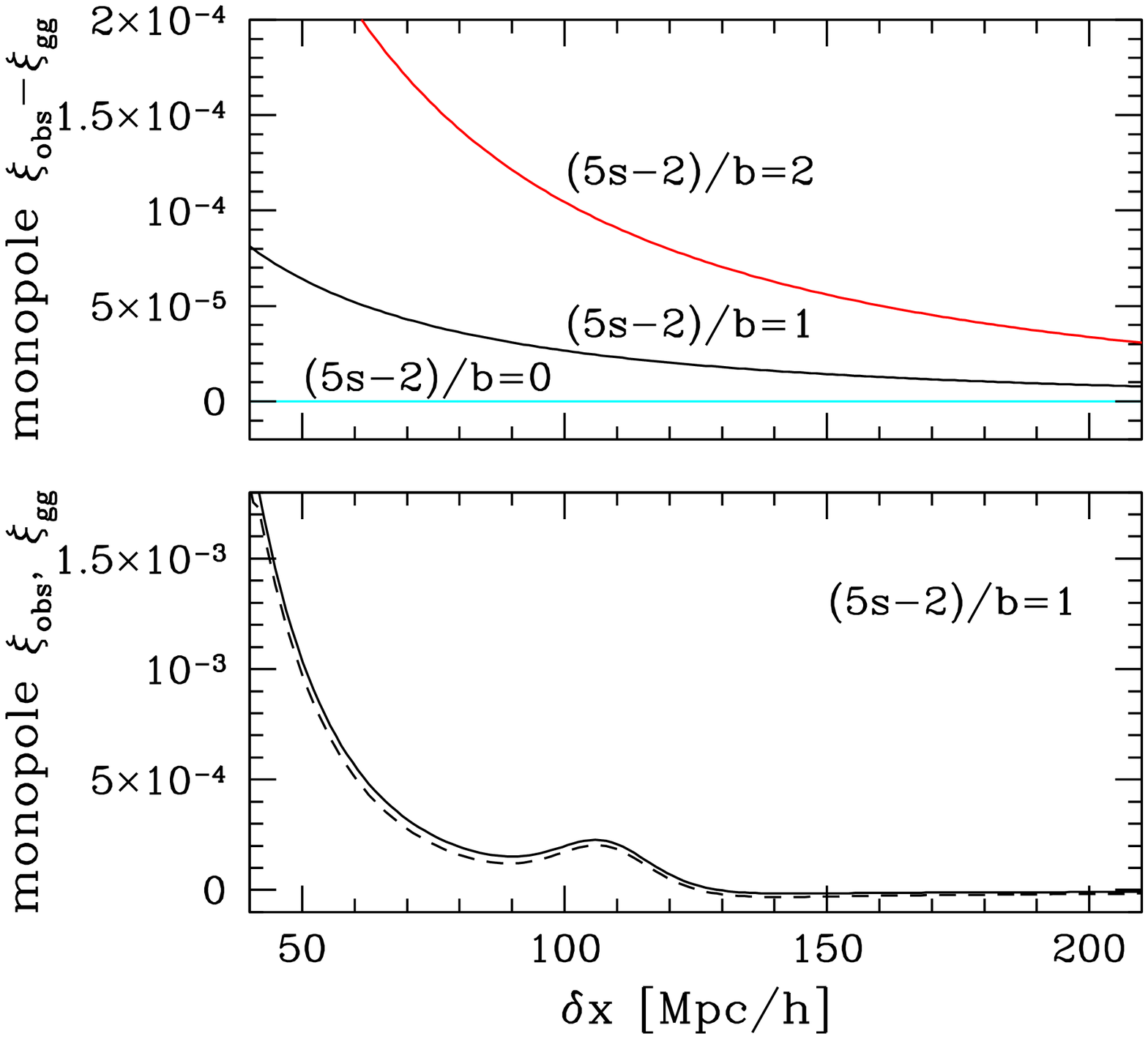}
}
\caption{Analog of Fig. \ref{xiall0.35} for $\bar z=3$. 
The contours in (a) are: 0.01 (magenta), 0.002 (cyan), 0.00045 (black),
0.00015 (blue; triple contours), 0 (red) and -0.00002 (green; double contours).
The dashed line in the lower panel
of ({\bf d}) is $\xi_{gg}$. 
}
\label{xiall3}
\end{figure*}

We can summarize the distinctive lensing induced anisotropy to the
observed correlation as follows:
\begin{eqnarray}
\label{scaling2}
\xi_{\rm obs} (\delta\chi, \delta x_\perp)
= && \xi_{gg} (\sqrt{\delta\chi^2 + \delta x_\perp^2}) \\ \nonumber && 
+ f(\delta x_\perp) \delta\chi
+ g(\delta x_\perp)
\end{eqnarray}
where $\delta\chi$ and $\delta x_\perp$ are the LOS and transverse separations
respectively, $f \delta\chi$ represents the galaxy-magnification correlation
and $g$ represents the magnification-magnification correlation.
Here, $f$ and $g$ are functions of the transverse separation only, and
are determined by the galaxy-mass and mass-mass power spectra.
{\it This distinctive form of the anisotropy allows us in principle
to separately measure $\xi_{gg}$, $f$ and $g$, from which we can
infer the galaxy-galaxy, galaxy-mass and mass-mass power spectra.}
For instance, at any given $\delta x_\perp$, plotting 
$\xi_{\rm obs}$ as a function of the LOS separation $\delta\chi$ would reveal a
linear contribution at sufficiently large $\delta\chi$'s
where $\xi_{gg}$ is very small. Its slope tells us
$f$ and its extrapolation to $\delta\chi=0$ tells us $g$. 
Subtracting $\delta\chi f + g$ from $\xi_{\rm obs}$ then yields $\xi_{gg}$.
This is illustrated in Fig. \ref{xi.fit}.
Fig. \ref{xi.fit} should be viewed as a proof of concept only.
The optimal method for extracting $\xi_{gg}, \xi_{g\mu}$ and $\xi_{\mu\mu}$ from
realistic data requires more investigation.

Let us study this distortion of the correlation function by magnification bias
in more quantitative detail.
Fig. \ref{xiall0.35} - \ref{xiall3} are a series of figures showing several
interesting aspects of the observed and intrinsic correlation functions for 
galaxies at mean redshifts of 
$\bar z = 0.35, 1, 1.5, 2, 3$. These redshifts are chosen to match roughly current
and future surveys \cite{baoexp}.
Here, as is throughout the paper, the correlation functions
shown in all figures are implicitly divided by $b^2$. 

Panels (a) show contours of constant $\xi_{gg}$ (dotted) and $\xi_{\rm obs}$ (solid)
as a function of the separation vector $\delta {\bf x} = {\bf x_1} - {\bf x_2}$.
Symmetry dictates that the two point correlations are functions of 
only the LOS projection $\delta\chi = |\chi_1 - \chi_2|$ (y-axis) and 
the transverse separation $\delta x_\perp = \bar\chi |\thetaB_1 - \thetaB_2|$ (x-axis).
Note how certain values of $\xi_{gg}$ or $\xi_{\rm obs}$ map to multiple contours:
this is because the correlation function is not monotonic, both around the baryon wiggle
and beyond the zero-crossing scale (see panels b).
Redshift distortion due to peculiar motion is ignored, and
therefore $\xi_{gg}$ is isotropic. 
The observed correlation $\xi_{\rm obs}$ is
anisotropic due to the presence of magnification bias corrections. These corrections
are most apparent for pair separations oriented along the LOS. In this orientation,
the galaxy-magnification correlation $\xi_{g\mu}$ grows linearly with separation (see eq. [\ref{gmu}]) 
while the intrinsic galaxy-galaxy correlation $\xi_{gg}$ drops with separation. 
At LOS separations $~70$ Mpc/h or larger, $\xi_{g\mu}$ dominates over $\xi_{gg}$.
As expected, the magnification distortion of the correlation function becomes increasingly
significant at larger redshifts -- the distortion is confined to small $\delta x_\perp$'s at low
redshifts, but diffuses to larger $\delta x_\perp$'s at high redshifts.
The overall distortion pattern vaguely resembles the well known finger-of-god (FOG) effect
due to virialized motion, in that the contours of constant $\xi_{\rm obs}$ are elongated
in the LOS direction. The precise shapes of the anisotropy, however, are rather different
- FOG corrections do not have this linear dependence on the LOS separation
that $\xi_{g\mu}$ has. Moreover, FOG due to peculiar
motion generally does not extend out to such large scales. We will examine the net clustering
anisotropy accounting for both magnification bias and peculiar motion
in Paper II. It is also worth emphasizing that $\xi_{g\mu}$ could have the opposite
sign if $s < 0.4$, in which case the LOS correlation is enhanced in the negative direction.
As can be seen from Fig. \ref{b.s3}, unless one has an exceptionally faint magnitude
cut-off, $s$ is generally larger than $0.4$ at high redshifts where gravitational lensing
is most effective.

Panels (b) show $\xi_{\rm obs}$ and its three different contributions for
a separation vector that is oriented along the LOS ($\theta_x = 0$, where
${\rm cos\,} \theta_x = \delta\chi/\delta x$ with $\delta x = |\delta {\bf x}| 
= \sqrt{\delta\chi^2 + \delta x_\perp^2}$). 
Eq. (\ref{mumu}) tells us $\xi_{\mu\mu}$ (flat dotted line) 
is independent of the LOS separation, and is therefore simply a constant for this 
particular orientation. As discussed earlier, $\xi_{g\mu}$ (sloped dotted line) 
grows linearly with separation in this orientation. The resulting
$\xi_{\rm obs} = \xi_{gg} + 2\xi_{g\mu} + \xi_{\mu\mu}$ is quite a bit larger
than $\xi_{gg}$, even at a redshift as low as $\bar z = 0.35$.
At a scale of $100$ Mpc/h, redshift $1.5$ appears to be the transition point
between galaxy-magnification dominance and magnification-magnification dominance
i.e. $2\xi_{g\mu} \gsim \xi_{\mu\mu}$ for $\bar z \lsim 1.5$,
and vice versa.
The inset shows the ratio $\xi_{\rm obs}/\xi_{gg}$
for several other orientations $\theta_x$ (solid lines). Comparing the insets
for the different redshifts, one can see that at low redshifts this ratio quickly drops
to unity as $\theta_x$ moves away from zero, while
the drop is much less rapid at high redshifts. This is consistent
with what we have seen in panel (a): the general diffusion of magnification 
distortion out to larger values of $\delta x_\perp$ (and therefore
$\theta_x$) as the redshift increases.
The dot-dashed line in the inset shows the monopole 
$\xi_{\rm obs}$ divided by $\xi_{gg}$. The monopole is defined to be
\begin{equation}
\label{monopole}
{\rm monopole \,\, of \,\, } \xi_{\rm obs} = 
\int_0^{\pi/2} \xi_{\rm obs} (\delta{\bf x}) {\,\rm sin}\theta_x d\theta_x \, .
\end{equation} 
At a scale of about $100$ Mpc/h, 
the fractional deviation of this monopole from $\xi_{gg}$ is negligible
at $\bar z = 0.35$, grows to a few percent at $\bar z = 1.5$ and is about $20 \%$
at $\bar z = 3$.

\begin{figure*}[tb]
\subfigure[Line-of-sight (LOS)]
{\label{shift.combo.LOS}\includegraphics[width=.45\textwidth]{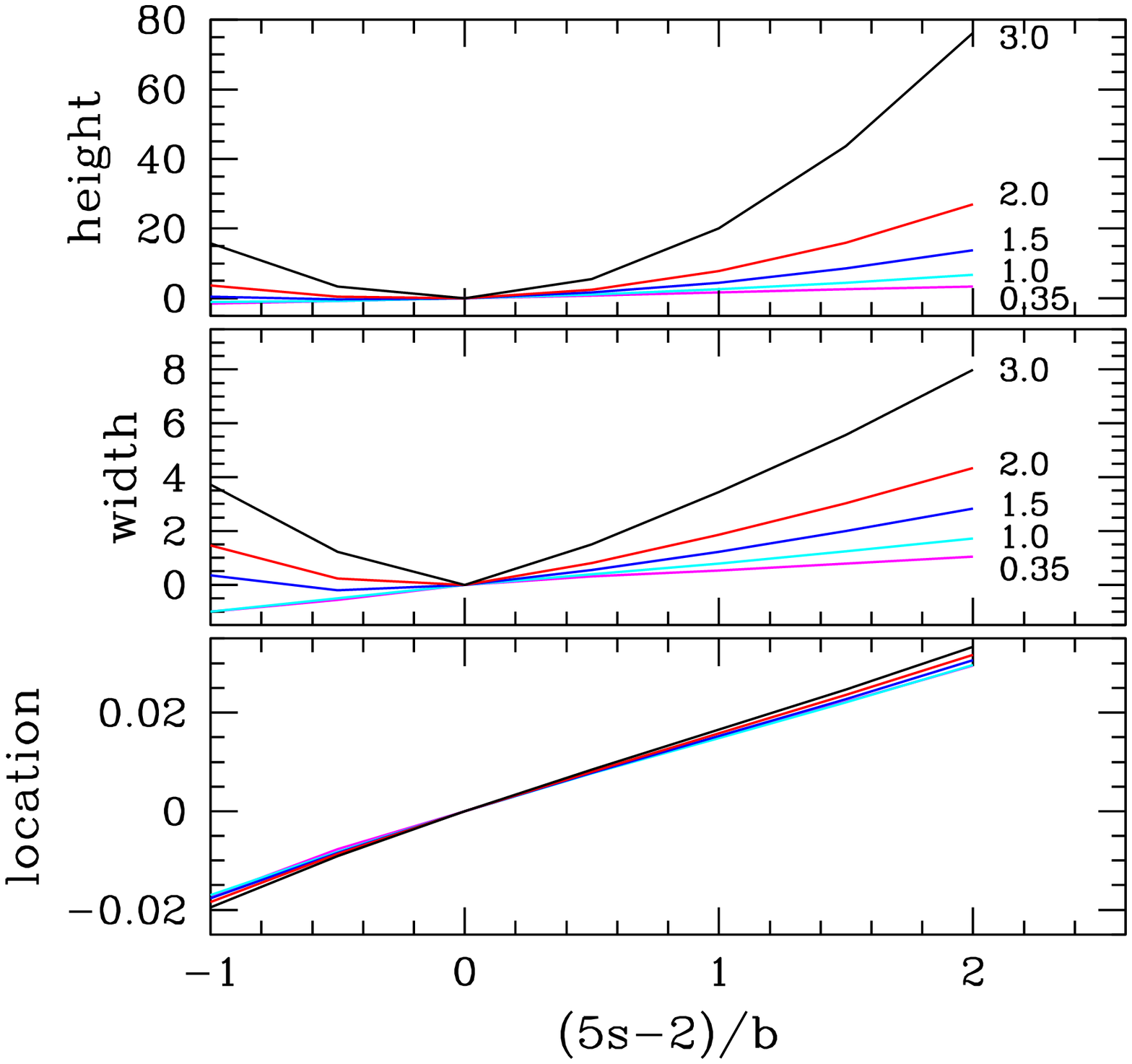}}
\hspace{0.1in}
\subfigure[Monopole]
{\label{shift.combo.monopole}\includegraphics[width=.45\textwidth]{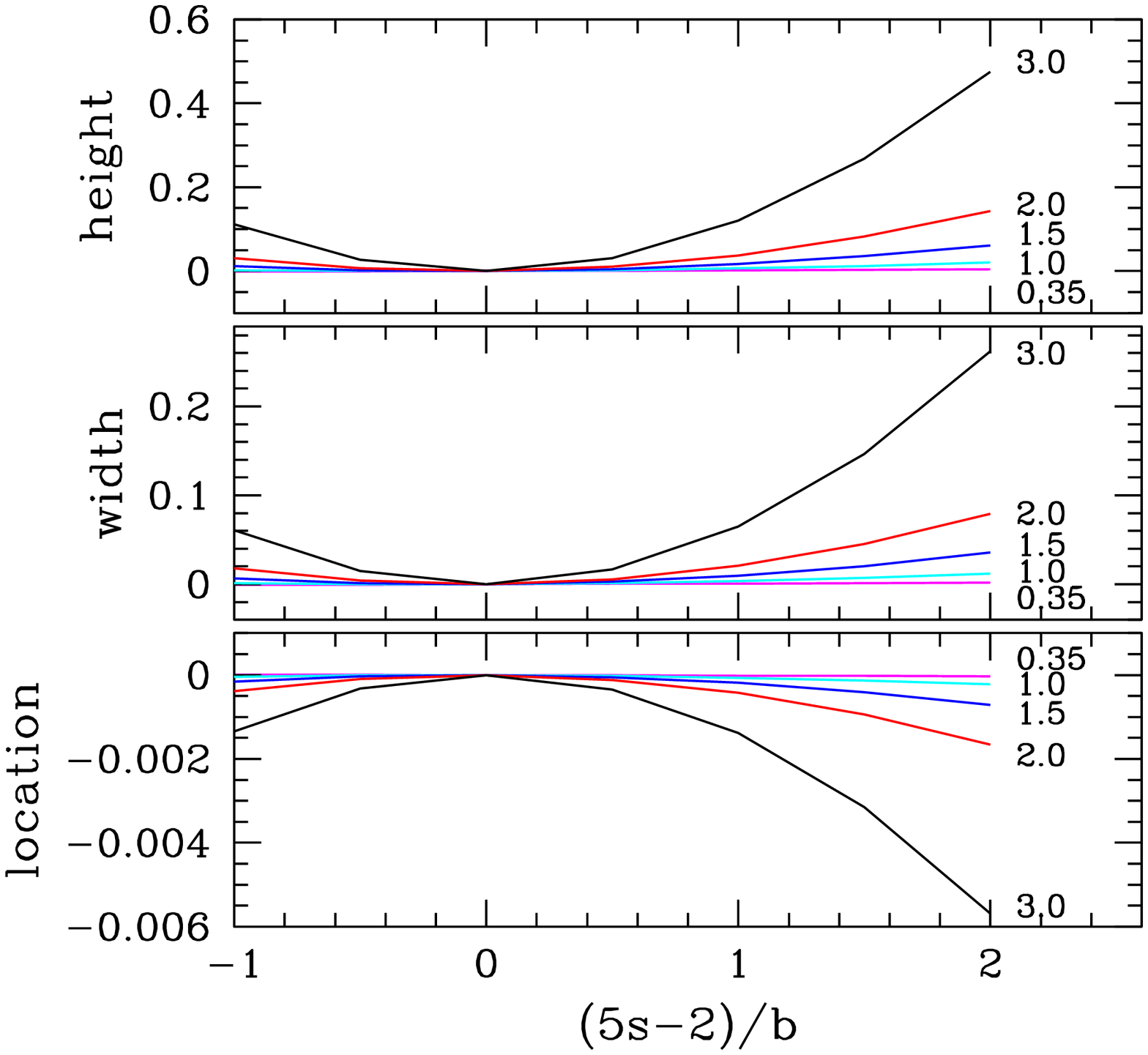}}
\caption{Fractional shifts in the baryon peak location, width and height as a function of
$(5s-2)/b$ for redshifts spanning $0.35, 1.0, 1.5, 2.0, 3.0$. 
Left panel is for the LOS orientation (where the change in peak location is very similar
for different redshifts and the redshift labels are therefore omitted), 
and right panel is for the monopole. The fractional shift is defined as
$(\xi_{\rm obs} - \xi_{gg}) / \xi_{gg}$ for the baryon peak height, and
similarly for the peak location and width. The peak width is defined as
$[d^2 \xi/d\delta x^2/\xi]^{-1/2}$ evaluated at the peak. A note of caution:
the precise shifts in these quantities depend on how they are extracted from
observations. Panels (d) of Fig. \ref{xiall0.35} - \ref{xiall3} provide a
more complete picture of the effects of magnification bias on the monopole.
}
\label{shift.combo}
\end{figure*}

\subsection{Baryon Acoustic Oscillations}
\label{sec:BAO}

Panels (c)  of Fig. \ref{xiall0.35} - \ref{xiall3} zoom in on the baryon wiggle for a separation vector
oriented along the LOS. Of interest, probably in descending order, are the location, width and height 
of the wiggle, and how much they are affected by magnification bias. The answers depend on
whether one looks at $\xi$ or, as is commonly done, $\delta x^2 \xi$.
An important lesson here is that $\delta x^2 \xi$
is in fact problematic: in the presence of magnification bias
(and for the LOS orientation), 
the baryon local maximum becomes difficult to locate or disappears completely!
Confining ourselves to $\xi$, the fractional changes in
the baryon peak location, width and height induced by magnification bias
are shown in the left panel of Fig. \ref{shift.combo} for the LOS orientation,
and for several different redshifts.
Here, we go beyond the assumption of $(5s-2)/b=1$ in most of the paper, and
show these fractional shifts for $(5s-2)/b$ spanning $-1$ to $2$. 
The shifts can be quite large. They can have either sign, depending in
part on the sign of $5s-2$. The magnitude of the shifts depends sensitively
on $(5s-2)/b$ and (for width and height) the redshift. All the shifts vanish
at the special point of $s=0.4$, where magnification bias is absent.
One surprise is that even at a low redshift of $0.35$, where one usually
expects lensing to be unimportant, the fractional shifts are significant: 
at $(5s-2)/b=1$, they are $\sim 2 \%$, $\sim 50\%$
and $\sim 170 \%$ for the location, width and height respectively!
This is fundamentally because the baryon wiggle is a large scale feature
$\sim 100$ Mpc/h. At such a large separation, the intrinsic galaxy-galaxy
clustering is quite weak. The lensing correction, on the other hand, is quite large
for the LOS orientation ($\theta_x = 0$) because one galaxy is directly behind another.
A location shift of $\sim 2 \%$ is significant for baryon oscillation
experiments \cite{baoexp}: 
it corresponds to a $\sim 2 \%$ shift in the inferred Hubble parameter
$H(z)$, which corresponds to a shift of
$\sim 10 \%$ for the dark energy equation of state $w$
at the redshifts of interest.

Fig. \ref{shift.combo} right panel shows
the shifts when one averages over
directions i.e. the monopole (eq. [\ref{monopole}]).
The shifts are much diminished, but in some cases, still non-negligible.
For redshifts $\lsim 1.5$, the location shift is $\lsim 0.1 \%$, 
but for redshifts $\gsim 1.5$, the location shift can reach up to $\sim 0.6 \%$.
The fractional shifts in width and height are, as before, a bit larger than
that in location. An intriguing aspect of Fig. \ref{shift.combo} is that,
comparing the LOS orientation and the monopole, the location shift actually
differs in sign for $(5s-2)/b > 0$. This suggests that the monopole is 
dominated by the transverse modes, which is not surprising, given that
the monopole downweighs the LOS orientation (eq. [\ref{monopole}]).

Note that the monopole can be inferred from data only in a cosmology dependent
manner i.e. assumptions need to be made in relating the 
observed redshift and angular separations
to separations in $\delta\chi$ and $\delta x_\perp$. 
Ignoring the anisotropy induced by magnification bias, it can be shown that
the baryon peak location in the monopole constrains the combination
$(\chi^2/H)^{1/3}$ (\cite{eisenstein}; see Appendix in Paper II). 
A $\sim 0.6 \%$ shift in this quantity translates into a $\sim 3 \%$ shift
in the dark energy equation of state $w$. 
One must keep in mind, however, that in the presence of anisotropy, such
as that induced by magnification bias, $(\chi^2/H)^{1/3}$ is 
not the exact quantity that is constrained by the monopole baryon peak.
We leave a thorough study of this issue for the future.

One must exercise additional caution in interpreting Fig. \ref{shift.combo}:
the precise shifts in the quantities shown depend on how they are
actually extracted from the observational data. Typically, one fits
some analytic curve through the data points to locate and characterize the
baryon peak. The important point is that magnification bias introduces
an additive correction to $\xi_{gg}$ which has a shape that is uncertain:
it depends on $b$, $s$ and the power spectrum. Igoring this correction
leads to errors in one's fit to the data, and therefore biases in
the inferred angular diameter distance and Hubble parameter. Precisely how
large a bias one would incur is procedure dependent, and deserves
a careful study.

Panels (d) of Fig. \ref{xiall0.35} - \ref{xiall3} emphasize this point by showing the difference: 
monopole of $\xi_{\rm obs} - \xi_{gg}$,
for several different values of $(5s-2)/b$. (We focus on the monopole instead
of the LOS orientation, because the monopole is less affected by magnification bias;
the difference $\xi_{\rm obs} - \xi_{gg}$ is quite a bit bigger for the LOS
orientation, see panels b of Fig. \ref{xiall0.35} - \ref{xiall3}.)
Comparing
against the monopole of $\xi_{\rm obs}$ itself, one can see that this difference
can be $\sim 10 \%$ already by $\bar z \sim 1.5$ at the baryon peak.
For a limited range of scales around the peak, one can perhaps
approximate this difference by some quadratic function, but the
precise slope and curvature of this quadratic are uncertain
and depend on the galaxy/quasar sample in question.
The issue is whether these should be treated as additional
parameters in one's fit and how they might affect the accuracy
of the peak location measurement. We leave this for a future paper.

\section{Interpretation}
\label{sec:interp}

Fig. \ref{xiall0.35} - \ref{shift.combo} are based on numerically evaluating
the integrals in eq. (\ref{gmu}), (\ref{mumu}) and (\ref{gg}). 
To gain a deeper understanding of what we have seen, 
it is useful to develop order of magnitude estimates for the ratios:
$\xi_{\mu\mu}/\xi_{gg}$ and $\xi_{g\mu}/\xi_{gg}$.
\begin{eqnarray}
\label{xiorder}
{\xi_{\mu\mu}\over \xi_{gg}} \sim \left[5s-2 \over b\right]^2 
{(1+ \bar z)^2\over 50}(\bar\chi H_0)^3 {\pi H_0\over k_*} {\Delta^2 (k_*) \over
\Delta^2 (k_{**})}  \\ \nonumber
{2\xi_{g\mu} \over \xi_{gg}} \sim  \left[5s-2 \over b\right] {1+\bar z \over 2}
(\delta\chi H_0) {\pi H_0\over k_*} {\Delta^2 (k_*) \over
\Delta^2 (k_{**})}
\end{eqnarray}
The symbol $\Delta^2 (k)$ denotes the
dimensionless variance at scale $k$ and redshift $\bar z$: $4\pi k^3 P_{mm} (k)/(2\pi)^3$. 
Here, $k_{**} \sim 1/\sqrt{\delta\chi^2 + \delta x_\perp^2}$,
while $k_*$ is equal to either $1/\delta x_\perp$ or $k_m$, whichever
is smaller ($k_m$ is the scale where $k^2 P_{mm} (k)$ peaks; $k_m \gsim 3$ h/Mpc).

The expressions in eq. (\ref{xiorder}) are only approximate. Among other things, they fail
at separations where the correlation functions cross zero or become negative.
Away from these separations, we have checked that these
expressions reproduce our numerical results quite well, to within factor of a few.
Moreover, they make transparent several important points.
There are two interesting limits to consider.

One is the limit of 
a small $\delta\chi$ and a large $\delta x_\perp$ ($\delta\chi \ll \delta x_\perp$), 
in which case the factors of $\Delta^2$'s cancel out. It is clear that both 
correlation ratios should then be small, as long as $\delta x_\perp$ ($\sim 1/k_*$) 
is a small fraction of the Hubble scale $H_0^{-1}$, and the redshift
is not too large.
A more intuitive way of putting it is that a transverse pair of galaxies
(i.e. $\delta \chi = 0$ and $\delta x_\perp \ne 0$)
are on average not significantly lensed: 
for one, they are not lensing each other i.e. the galaxy-magnification term
vanishes; as far as the magnification-magnification term is concerned,
it is generally small compared to the intrinsic galaxy correlation 
unless the redshift is sufficiently high. 

The other interesting limit is that of a large $\delta\chi$ and a small 
$\delta x_\perp$ ($\delta\chi \gg \delta x_\perp$),
in which case $\Delta^2 (k_*) \gg \Delta^2 (k_{**})$
(because $k_* \sim 1/\delta x_\perp$ and
$k_{**} \sim 1/\delta\chi$). This is the limit
in which the correlation ratios in eq. (\ref{xiorder}) can potentially be 
of order unity or even larger,
with the help of the large boost from $\Delta^2 (k_*) / \Delta^2 (k_{**}) \gg 1$.
What is the physical origin of this boost?
Consider a pair of galaxies oriented along the LOS (i.e. $\delta \chi \ne 0$ and $\delta x_\perp = 0$).
When $\delta\chi$ is large, the intrinsic galaxy correlation is quite weak.
The magnification bias corrections, on the other hand, can be relatively large
because this is the case of zero lensing impact parameter ($\delta x_\perp = 0$). 
In other words, one is fundamentally comparing
fluctuations on very different scales: large or linear scales in the case of
the intrinsic galaxy clustering $\xi_{gg}$,
and small or nonlinear scales in the case of the lensing corrections $\xi_{g\mu}$
and $\xi_{\mu\mu}$. The naive expectation is that 
one is performing clustering measurements on linear scales by considering
pairs of galaxies separated at a large $\delta\chi$ (say $100$ Mpc/h), but the truth
is that the observed clustering is dominated by nonlinear fluctuations
due to lensing. 

Incidentally, this also means our linear galaxy bias assumption
is probably not a good approximation for pair separations oriented along, or
close to, the LOS. The galaxy bias is generally expected to first drop
below the linear value for scales just below the nonlinear scale, and then
climb when approaching zero lag \cite{JSS03}. 
This means the effects of gravitational lensing on the LOS orientation should
be {\it enhanced} by a nonlinear galaxy bias compared to our predictions.
One can see this by recalling the expression for the observed correlation:
$\xi_{\rm obs} = \xi_{gg} + 2 \xi_{g\mu} + \xi_{\mu\mu}$.
In the LOS orientation, and for a large separation, the relevant bias
for $\xi_{gg}$ is the linear galaxy bias $b$ that we have been using all along,
while the relevant bias for $\xi_{g\mu}$ (at $\delta x_\perp = 0$) 
should really be a nonlinear and presumably enhanced galaxy bias $> b$.
Therefore, the effects of gravitational lensing are in fact {\it underestimated}
by our calculations which use a linear galaxy bias.
The precise scale dependence though of the nonlinear galaxy bias
is quite complicated and is highly sample dependent. We hope to pursue
a thorough study of the effects of nonlinear galaxy bias in the future.
It is important to stress that the distinctive anisotropy displayed in
eq. (\ref{scaling2}) holds true even in the presence of nonlinear galaxy bias
-- this is because eq. ({\ref{scaling2}) follows from eq. (\ref{gmu}),
(\ref{mumu}) and (\ref{gg}) which make no assumptions about galaxy biasing.

Irrespective of details of the nonlinear galaxy bias, the following
statement is expected to hold: 
{\it magnification-bias has the strongest
effect on the observed correlation
function when the separation $\delta{\bf x}$ has a large
magnitude and points close to the line-of-sight direction.}

Eq. (\ref{xiorder}) also tells us that $\xi_{\mu\mu}$ is generally
expected to be larger than $\xi_{g\mu}$ when the redshift is sufficiently
large. Sufficiently large here means when $\bar\chi H_0 \gsim 1$. 
This occurs at $\bar z \sim 1.5$ for the cosmological parameters we adopt.
Conversely, $\xi_{\mu\mu}$ is generally quite small at low redshifts because
of the cubic power in the factor $(\bar\chi H_0)^3$. In this case,
$\xi_{g\mu}$ is more favored if the LOS separation $\delta\chi$ is large enough.

\section{Discussion}
\label{discuss}

Our findings can be summarized as an opportunity and a challenge.

{\it Opportunity.} Gravitational lensing, through magnification bias,
introduces a distinctive anisotropy to the observed 3D galaxy/quasar correlation
function (eq. [\ref{scaling2}]). We have studied its shape in
\S \ref{correlation}.
The correlation is preferentially enhanced (in the positive
direction for $s > 0.4$) in the LOS orientation, vaguely
resembling the finger-of-god (FOG) effect due to virialized peculiar motions.
However, as will be discussed in detail in Paper II, the precise shape
of magnification distortion differs from that of FOG or redshift distortion
in general -- redshift distortion does not have the distinctive linear
dependence on the LOS separation that $\xi_{g\mu}$ has.
The distinctive lensing induced pattern creates an interesting opportunity: 
it is in principle possible to separately obtain
from data all three contributions: the galaxy-galaxy, galaxy-magnification
{\it and} magnification-magnification correlations (Fig. \ref{xi.fit}).
This generalizes earlier work on magnification bias \cite{MJ98,EGmag03,ScrantonSDSS05,menard,bhuv} 
which focused on angular correlations in the large redshift separation limit,
where only the galaxy-magnification correlation survives.
Future galaxy/quasar (spectroscopic or photometric) redshift surveys can be used
to measure the galaxy-galaxy, galaxy-mass and mass-mass power spectra, even without
measurements of the galaxy shapes.

{\it Challenge.} The same effect must be accounted for in interpreting 
galaxy/quasar clustering data. This presents an interesting challenge to
precision measurements, such as those that hope to use the baryon oscillation
scale to yield stringent constraints on dark energy.
Contrary to naive expectations, magnification bias is not necessarily negligible
at low redshifts. In the LOS orientation, where its effects are largest,
the shift in the baryon oscillation scale is remarkably insensitive to redshifts,
from $\bar z \sim 0.35$ to $\bar z \sim 3$ (Fig. \ref{shift.combo} panel a).
The shift could reach up to $\sim 2 - 3 \%$, which translates to a
$\sim 10 - 15 \%$ shift in the dark energy equation of state from
the inferred Hubble parameter.
Existing baryon oscillation measurements \cite{eisenstein,2dFa,2dFb,hutsi,tegmark} have not
reached this kind of accuracy, but the future ones will \cite{baoexp}.
The shift for other orientations are smaller. We have considered the
shift in the monopole and find that it is much diminished, reaching
up to $\sim 0.6 \%$ at $\bar z \sim 3$ (Fig. \ref{shift.combo} panel b).
However, one should keep in mind that the precise shifts really depend on
exactly how the baryon oscillation scale is extracted from data.
The point is that magnification bias introduces corrections to the
correlation function that are dependent on scale, bias and the number
count slope, for all orientations and for the monopole (Fig. \ref{xiall0.35}
- \ref{xiall3} panels d).
The question is when data are fitted without accounting for these corrections,
what kind of bias would one incur in the inferred parameters?

This leads us naturally to several interesting questions to
be explored in the future.

On the opportunity side, what is the optimal scheme to extract the galaxy-galaxy,
galaxy-magnification and magnification-magnification correlations from
realistic data? Fig. \ref{xi.fit} represents a proof of concept but 
is likely not the optimal method. 
How to best weigh the relative contributions of different galaxies with different
bias and number count slope?
Also, how to best guard against the possibility of confusion with dust extinction?
The distinctive dependence of the magnification corrections on the number count
slope can probably be exploited for this purpose \cite{ScrantonSDSS05}.

On the challenge side, how would magnification bias impact baryon oscillation
measurements if one accounts for precisely how the oscillation scale might
be extracted from future data? How might the impact be different in Fourier space
versus real space? We will address this question partially in Paper II.

Our calculations assumed a constant bias $b$ which is certainly incorrect at some level.
A complete calculation that accounts for nonlinear bias and distortions
of all types (peculiar motions, Alcock-Paczynski and magnification bias)
should be made to properly interpet precision galaxy clustering measurements
\cite{nishimichi}.
Lastly, in light of our findings, it is probably interesting to revisit some
longstanding puzzles, such as the well known results of the pencil beam surveys by
\cite{pencil,szalay}. The enhancement of the LOS correlation by magnification bias
offers a plausible explanation for the excess correlations, though not
the periodicity, seen in \cite{pencil}.
With the magnitude cut-off used by \cite{pencil},
it is likely that $(5s - 2)/b \sim 1$ (see also \cite{metcalfe}), 
in which case we can scale the result of Fig. \ref{xiall0.35}(c) to 
the mean redshift of $\sim 0.2$ using eq. (\ref{xiorder}) and obtain an
enhancement factor of $\sim 2$ at the scale $\sim 100$ Mpc/h.
Coupled this with the fact that the baryon wiggle was not taken into
consideration back in the days of \cite{pencil}, the total enhancement
factor by magnification bias $+$ baryon oscillations is $\sim 4$.
Moreover, as noted in \S \ref{sec:interp}, accounting for nonlinear
galaxy bias in the lensing correction would likely lead to a further enhancement
factor of a few. The net enhancement factor could well be $\sim 10$, making
the excess correlations seen in \cite{pencil} within reach.
It would be worthwile to carry out a more careful analysis, taking into account
properties of the precise galaxy sample in question, and the actual widths of the 
pencil beams.
It might also be interesting to revisit
some of the measurements of Lyman-break galaxy clustering at $z \sim 3$ \cite{LBGs} 
in the light of our findings.

\acknowledgments

We thank Roman Scoccimarro and Albert Stebbins for useful discussions.
We also thank Taka Matsubara and Asantha Cooray for pointing out to us an
important missing reference in our first preprint.
Research for this work is supported by the DOE, grant DE-FG02-92-ER40699,
and the Initiatives in Science and Engineering Program
at Columbia University. EG acknowledges support from Spanish Ministerio de Ciencia y
Tecnologia (MEC), project AYA2006-06341 with EC-FEDER funding, and
research project 2005SGR00728 from  Generalitat de Catalunya.

\appendix

\section{Magnification Bias and Other Lensing Corrections to the Observed Galaxy Clustering}
\label{app:lens}

In this Appendix, we attempt to be as general as possible in considering the effect
of gravitational lensing on the apparent spatial clustering of galaxies.
In the process, we will rederive the magnification bias effect, generalizing it to
the case of a gradual rather than sharp magnitude cut-off. We will also discuss the role of
stochastic deflections which contribute to higher order corrections.

Let $\Phi_g (f_g, z, \thetagB) df_g$ be the intrinsic number density of galaxies 
at redshift $z$, source (unlensed) position $\thetagB$, and an unlensed 
flux in the range $f_g \pm df_g/2$. 
We use the subscript $g$ to denote quantities that are intrinsic
to the galaxies, or in other words, prelensed/unlensed quantities.
Note also that the number density here can refer to either an angular number density
or a volume number density. 
Gravitational lensing introduces the transformations
$f_g \rightarrow f$, $\thetagB \rightarrow \thetaB$ and $\Phi_g \rightarrow \Phi$, which
satisfy the following relations:
\begin{eqnarray}
\label{Phi}
\Phi (f, z, \thetaB) df d^2\theta = \Phi_g (f_g, z, \thetagB) df_g d^2\theta_g
\end{eqnarray}
where
\begin{eqnarray}
\label{transf}
\thetagB = \thetaB + \dthetaB \\ \nonumber
f = A f_g \\ \nonumber
{\rm det.} \left[{\partial \theta_g^i \over \partial \theta^j}\right] \equiv 1/A
\end{eqnarray}
The quantity $A$ denotes the magnification in the observed direction $\thetaB$, 
and $\dthetaB$ is the lensing displacement.

A given galaxy sample can be modeled by an efficiency function $\epsilon(f)$
such that the observed galaxy density is 
\begin{eqnarray}
\label{epsilon}
n(z, \thetaB) = \int \epsilon(f) \Phi (f, z, \thetaB) df
\end{eqnarray}
The simplest example of $\epsilon(f)$ is a step function which 
equals unity if $f > f_{\rm min.}$ and vanishes otherwise:
\begin{eqnarray}
\label{Theta}
\epsilon(f) = \Theta (f_{\rm min.} < f)
\end{eqnarray}
but let us keep $\epsilon(f)$ general.

Using eq. (\ref{Phi}) and (\ref{transf}), we can see that
\begin{eqnarray}
\label{nzero}
n(z, \thetaB) = {1 \over A(z, \thetaB)} \int \epsilon(A f_g) \Phi_g (f_g, z, \thetaB + \dthetaB) df_g 
\end{eqnarray}
Note that the lensing magnification $A$ is a function of both the source redshift $z$ and
the direction $\thetaB$. 

Our expressions so far are completely general: no weak lensing approximation has been
made (aside from not summing over multiple $\dthetaB$'s to account for the possibility
of multiple images). One could if one wishes use eq. (\ref{nzero}) as the starting point for
investigating the lensed galaxy clustering. 
The discussion is much simplified, however, under the weak lensing approximation i.e.
$A \sim 1 + 2\kappa$, where $|\kappa| \ll 1$. 
In that case,
\begin{eqnarray}
n(z, \thetaB) = n_g (z, \thetaB + \dthetaB) \left[ 1 + (5s - 2) \kappa \right]
\end{eqnarray}
where 
\begin{eqnarray}
n_g = \int \epsilon(f_g) \Phi_g (f_g) df_g
\end{eqnarray}
is the galaxy density if lensing magnification were absent (we have suppressed the
$z$ and $\thetaB + \dthetaB$ dependence),
and
\begin{eqnarray}
\label{sdef}
s \equiv [2.5 n_g]^{-1}
\int {d\epsilon \over df_g} f_g \Phi_g (f_g) d f_g
\end{eqnarray}
The definition of $s$ might look a bit unfamiliar, but if $\epsilon$ does indeed
take the form of the step function (eq. [\ref{Theta}]), $s$ is the (magnitude) slope of the
cumulative number counts at the faint end cut-off:
\begin{eqnarray}
s = {d {\rm log}_{10} n_g \over d m}
\end{eqnarray}
where $m \equiv -2.5 \,{\rm log}_{10} f$ and the derivative is evaluated at $m = -2.5 \, {\rm log}_{10} f_{\rm min.}$. 

As long as $\langle A \rangle = 1$ (or $\langle \kappa \rangle = 0$) which holds
if multiple imaging can be ignored, $\langle n \rangle = \langle n_g \rangle$. 
Let us call this mean density $\bar n$. 
The fluctuation in the observed galaxy density $\delta \equiv n/\bar n - 1$ is
related to the fluctuation in the intrinsic galaxy density $\delta_g \equiv n_g/\bar n - 1$ by
\begin{eqnarray}
\label{complete}
&& \delta(z, \thetaB) = \delta_g(z, \thetaB + \dthetaB) + \\ \nonumber
&& \quad (5s - 2)\kappa (z, \thetaB) \times
(1 + \delta_g(z, \thetaB + \dthetaB))
\end{eqnarray}

This is equivalent to eq. (\ref{start}) and (\ref{deltamu}) used in the rest of the paper,
except that we have consistently ignored the effects of stochastic deflection $\dthetaB$,
and we have consistently neglected $\kappa \delta_g$ as small.
In other words, gravitational lensing strictly speaking should have 2 qualitatively different
effects on the observed galaxy clustering. Magnification bias is embodied in
the term $(5s-2) \kappa$, which accounts for the effect of an overall magnification
(or demagnification) on the number counts. Stochastic deflections, on the other hand, introduces
a remapping of the galaxy density field. The remapped field can be Taylor expanded as
\begin{eqnarray}
\delta_g (\thetaB + \dthetaB) \sim \delta_g (\thetaB) + \sum_i {\partial \delta_g \over \partial \theta_i} \delta\theta_i
\end{eqnarray}
The approximation we have been making is in effect a linear approximation: the second term 
on the right is second order in perturbations and is therefore ignored
(just as we throw away the $\kappa \delta_g$ term also).
Similarly, the $s$ defined in eq. (\ref{sdef}) can be replaced by
one with $f_g \rightarrow f$ and $\Phi_g \rightarrow \Phi$. Corrections to this
when multiplied by $\kappa$ as in eq. (\ref{complete}) are second order.

It is worth emphasizing that the gravitational lensing effect on diffuse backgrounds, such
as the microwave background or redshifted 21cm background, is
quite different \cite{stochastic,zahn}. There, because the observable is specific intensity which is 
conserved by lensing, the {\it only} lensing effect is remapping by stochastic deflections i.e. there is no
analog of the magnification bias term $(5s-2)\kappa$. 

One might worry that in the nonlinear regime
where $\delta_g$ is not small, one should not ignore the effects of
stochastic deflection relative to magnification bias. 
This certainly deserves more study.
Zahn \& Zaldarriaga \cite{zahn} carried out a detailed study of
the impact of stochastic deflection on the redshifted
21cm background, including all higher order terms. Stochastic deflection
appears to have a rather small overall impact, suggesting that
ignoring it might not be a bad approximation. 

\section{Higher Order Corrections}
\label{app:lens2}

Eq. (\ref{gmu}), (\ref{mumu}) and (\ref{gg}), and the corresponding
expressions in Fourier space, are written down by keeping only the dominant
terms in an expansion using $(\chi_1 - \bar\chi)/\bar\chi$ or $(\chi_2 - \bar\chi)/\bar\chi$
as a small parameter. It is worth discussing how large the corrections are expected to be.
One might naively expect the corrections to be one order higher:
for instance, that the dominant correction to $\xi_{\mu\mu}(1;2)$ should be of the order of
what is given in eq. (\ref{mumu}) multiplied by $(\chi_1 - \chi_2)/\bar\chi$. 
This expectation turns out to be false for $\xi_{\mu\mu}$ and $\xi_{gg}$, but valid for $\xi_{g\mu}$.
Let us see how this comes about in the case of $\xi_{\mu\mu}$. Taylor expanding eq. (\ref{mumuexact})
for $\xi_{\mu\mu}(1;2)$
in small $\delta\chi_1 = \chi_1 - \bar\chi$ and $\delta\chi_2 = \chi_2 - \bar\chi$, we have
\begin{eqnarray}
\xi_{\mu\mu} (\chi_1, \thetaB_1; \chi_2, \thetaB_2) =
[{3\over 2} H_0^2 \Omega_m (5s - 2)]^2 \\ \nonumber
\int_0^{\bar\chi} d\chi' \left[{(\bar\chi - \chi')\chi' \over \bar\chi}\right]^2 (1+z')^2 \\ \nonumber
\int {d^2 k_\perp \over (2\pi)^2} P_{mm} (z', k_\perp) e^{i {\bf k_\perp} \cdot \chi' (\thetaB_1 - \thetaB_2)} \\ \nonumber
[1 + O(\delta\chi_1/\bar\chi) + O(\delta\chi_2/\bar\chi)]
\end{eqnarray}
where the terms $O(\delta\chi_1/\bar\chi)$ and $O(\delta\chi_2/\bar\chi)$ signify all the next-to-dominant
order corrections which in general do not vanish. The important realization is that the
term $O(\delta\chi_1/\bar\chi)$ and the term $O(\delta\chi_2/\bar\chi)$ have exactly the same coefficients,
and so they sum to zero as long as $\delta\chi_1 + \delta\chi_2 = 0$ i.e.
$\bar\chi$ is the mean of $\chi_1$ and $\chi_2$.
Therefore, one has to go one order higher in the Taylor expansion to figure out
the actual corrections. 
A similar argument works for the galaxy auto-correlation $\xi_{gg}$. The
magnification-galaxy cross-correlation $\xi_{g\mu}(1;2)+\xi_{g\mu}(2;1)$, 
on the other hand,
turns out to be different: in this case, the analogous terms 
$O(\delta\chi_1/\bar\chi)$ and $O(\delta\chi_2/\bar\chi)$ have different coefficients.
In general, when the pair $(1;2)$ is averaged over some survey volume, one expects
the (fractional) correction to our expression for the
magnification-galaxy cross-correlation to be $O(|\chi_1 - \chi_2|/\bar\chi)$. 
If one is worried about this correction, there is no reason why one cannot
use instead the exact expression (within Limber approximation) for $\xi_{g\mu}$ in eq. (\ref{gmuexact}).

\newcommand\spr[3]{{\it Physics Reports} {\bf #1}, #2 (#3)}
\newcommand\sapj[3]{ {\it Astrophys. J.} {\bf #1}, #2 (#3) }
\newcommand\sapjs[3]{ {\it Astrophys. J. Suppl.} {\bf #1}, #2 (#3) }
\newcommand\sprd[3]{ {\it Phys. Rev. D} {\bf #1}, #2 (#3) }
\newcommand\sprl[3]{ {\it Phys. Rev. Letters} {\bf #1}, #2 (#3) }
\newcommand\np[3]{ {\it Nucl.~Phys. B} {\bf #1}, #2 (#3) }
\newcommand\smnras[3]{{\it Monthly Notices of Royal
        Astronomical Society} {\bf #1}, #2 (#3)}
\newcommand\splb[3]{{\it Physics Letters} {\bf B#1}, #2 (#3)}

\newcommand\AaA{Astron. \& Astrophys.~}
\newcommand\apjs{Astrophys. J. Suppl.}
\newcommand\aj{Astron. J.}
\newcommand\mnras{Mon. Not. R. Astron. Soc.~}
\newcommand\apjl{Astrophys. J. Lett.~}
\newcommand\etal{{~et al.~}}

\end{document}